\documentclass[twocolumn,prd,superscriptaddress,nofootinbib,preprintnumbers]{revtex4}

\usepackage{graphicx}
\usepackage{amsmath,bm,amssymb,amsfonts,dsfont}
\usepackage[usenames,dvipsnames]{xcolor}
\usepackage[normalem]{ulem}
\usepackage{url}
\usepackage{multirow}
\usepackage[colorlinks  = true,
            linkcolor   = NavyBlue,
            urlcolor    = NavyBlue,
            citecolor   = NavyBlue,
            anchorcolor = NavyBlue]{hyperref}

\pdfoutput=1 

\newcommand{\lsim}{\mathrel{\mathop{\kern 0pt \rlap
  {\raise.2ex\hbox{$<$}}}
  \lower.9ex\hbox{\kern-.190em $\sim$}}}
\newcommand{\gsim}{\mathrel{\mathop{\kern 0pt \rlap
  {\raise.2ex\hbox{$>$}}}
  \lower.9ex\hbox{\kern-.190em $\sim$}}}

\def \aap  {Astronomy \& Astrophysics}

\def \apj  {Astrophys. J.}

\def \mnras {MNRAS}

\newcommand{\db}{\texttt{diff.brk}}
\newcommand{\dbin}{\texttt{diff.brk inhom}}
\newcommand{\ib}{\texttt{inj.brk$+v_{\rm A}$}}

\begin{document}

\title{
Data-Driven Constraints on Cosmic-Ray Diffusion: Probing Self-Generated Turbulence in the Milky Way
}

\author{Mattia Di Mauro}
\email{dimauro.mattia@gmail.com}
\affiliation{Istituto Nazionale di Fisica Nucleare, via P. Giuria, 1, 10125 Torino, Italy}

\author{Michael Korsmeier}
\email{michael.korsmeier@fysik.su.se}
\affiliation{Stockholm University and The Oskar Klein Centre for Cosmoparticle Physics, Alba Nova, 10691 Stockholm, Sweden}

\author{Alessandro Cuoco}
\email{alessandro.cuoco@unito.it}
\affiliation{Department of Physics, University of Torino, via P. Giuria, 1, 10125 Torino, Italy}
\affiliation{Istituto Nazionale di Fisica Nucleare, via P. Giuria, 1, 10125 Torino, Italy}

\begin{abstract}
We employ a data-driven approach to investigate the rigidity and spatial dependence of the diffusion of cosmic rays in the turbulent magnetic field of the Milky Way. Our analysis combines data sets from the experiments Voyager, AMS-02, CALET, and DAMPE for a range of cosmic ray nuclei from protons to oxygen. Our findings favor models with a smooth behavior in the diffusion coefficient, indicating a good qualitative agreement with the predictions of self-generated magnetic turbulence models. Instead, the current cosmic-ray data do not exhibit a clear preference for or against inhomogeneous diffusion, which is also a prediction of these models. Future progress might be possible by combining cosmic-ray data with gamma rays or radio observations, enabling a more comprehensive exploration.
\end{abstract}

\maketitle

\section{Introduction}
\label{sec:intro}

The Alpha Magnetic Spectrometer (AMS-02) on the International Space Station has achieved remarkable precision in measuring the flux of various cosmic-ray (CR) nuclei from protons to iron, with unprecedented accuracy at the level of a few percent \cite{AMS:2021nhj}. By studying these CR nuclei, we can deduce valuable insights into CR propagation and gain a deeper understanding of the magnetic fields and turbulence of our Galaxy. 
The increase in precision and the large range of measured nuclei and isotopes allow for testing increasingly sophisticated models with higher levels of refinement.

The combination of primary and secondary cosmic rays, such as the boron-to-carbon (B/C) ratio, plays a crucial role in determining the grammage of CRs, i.e., the average amount of gas traversed by CRs during their journey. Additionally, radioactive isotopes like $^{10}$Be serve as clocks, enabling us to estimate the average propagation time of CRs before they reach our detectors (see e.g. \cite{Evoli:2017vim,Weinrich:2020ftb,DiMauro:2021qcf,Korsmeier:2021bkw,Korsmeier:2021brc,Maurin:2022gfm,Genolini:2021doh}).
For instance, on average, a CR nucleus with an energy of 10 GeV encounters a grammage of approximately 10~$\mathrm{g/cm^2}$  and propagates for about $\mathcal{O}(10^6)$ years \cite{Blasi:2013rva}.

These observations suggest that CRs diffuse within a region much larger than the gaseous Galactic disc, extending only a couple hundred pc above and below the Galactic plane. In contrast, the diffusion region, commonly called the halo, extends over kpc scales. The exact determination of the size $L$ of the diffusion halo is challenging, but recent analyses based on Be data from AMS-02 suggest that it exceeds a few kpc \cite{Evoli:2019iih,Cuoco:2019kuu,Maurin:2022gfm,DiMauro:2023oqx}.

Independent evidence for the existence of a magnetized halo of several kpc comes from radio observations, revealing the presence of electrons and magnetic fields producing synchrotron emission \cite{Bringmann:2011py,DiBernardo:2012zu,Orlando:2013ysa} above and below the Galactic plane. 
Furthermore, investigations of the diffuse $\gamma$-ray background \cite{2012ApJ...750....3A} support the existence of an extended halo.

\begin{figure}
\centering
\includegraphics[width=0.49\textwidth]{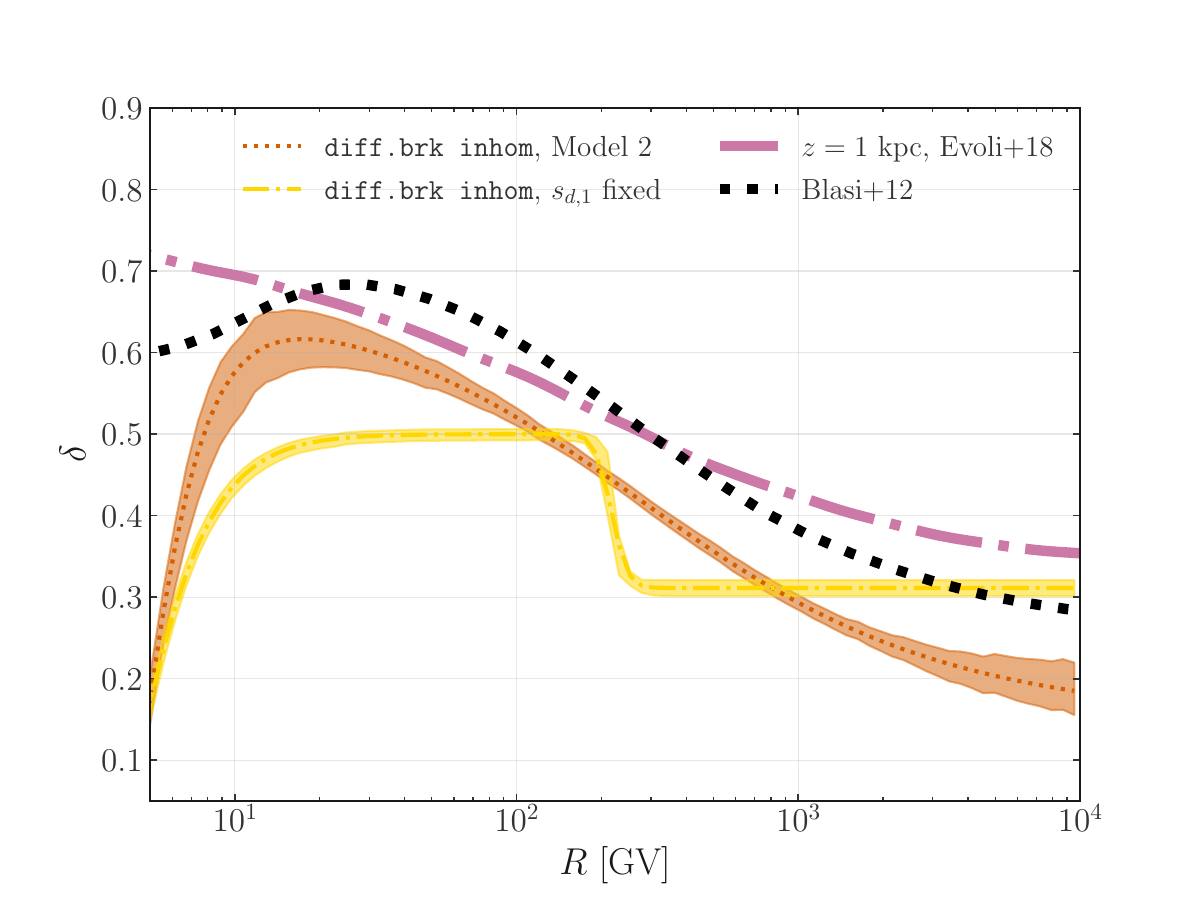}
\caption{Slope of the CR diffusion coefficient as a function of rigidity (calculated as $\delta = dD/dR$). CR data from AMS-02, CALET, and DAMPE significantly prefer a smooth transition (orange band) of the diffusion coefficient compared to a sharp transition of the power law (yellow line). Such a smooth transition is predicted in models of self-generated magnetic turbulence in the Galaxy \cite{Blasi:2012yr,Evoli:2018nmb} (black and magenta dashed lines).}
\label{fig:main_result}
\end{figure}

The prevailing understanding suggests that primary CR nuclei originate from and are accelerated by supernova remnants (SNRs) through a process known as diffusive shock acceleration \cite{1977DoSSR.234.1306K,10.1093/mnras/182.2.147}. Following their production,  primary CRs propagate within the Galactic environment, where they diffuse through the turbulent magnetic field, interact with interstellar gas, and may also be affected by Galactic winds. In certain catastrophic interactions, CRs can fragment, giving rise to secondary CRs. Species such as Li, Be, and B are predominantly composed of secondary CRs, while p, He, as well as C, N, and O are mostly of primary origin. Before CRs reach our detectors, they traverse the Heliosphere, where they are deflected and decelerated by solar winds, a process known as solar modulation \cite{Fisk:1976aw}.

The propagation of CRs is typically described by a phenomenological model proposed in Ref.~\cite{Ginzburg,1990acr..book.....B}. The model  assumes that magnetic turbulence is injected into the system by supernovae at large scales. Over time, the energy spectrum of the turbulence cascades to smaller scales in accordance with theories of turbulence. Consequently, this results in a power-law distribution of the wave power spectrum for the magnetic turbulence, and subsequently, of the diffusion coefficient. To perform practical calculations, it is often assumed that the diffusion coefficient remains homogeneous and isotropic within a cylindrical halo surrounding the Galaxy. Beyond this magnetic halo, the turbulence level is assumed to diminish, allowing particles to escape into intergalactic space freely.

The model and assumptions outlined above, although valuable as a first-order approximation, do not incorporate feedback from CRs on the magnetic turbulence spectrum. While this simplification is reasonable to some extent, recent high-precision data from the AMS-02 experiment allow us to challenge this notion. Indeed, the observed spectral break in CR nuclei at rigidities between 200 and 300 GV provides evidence for a deviation from the simple power law scenario. This break is more prominent in secondary CRs compared to primary CRs \cite{AMS:2018tbl}, suggesting a change in the slope of the diffusion coefficient as the likely explanation \cite{Genolini:2017dfb}. 

Indeed, the presence of a gradient in CR density can lead to the production of self-generated turbulence, a process that has gained attention in various astrophysical settings. For instance, in the vicinity of pulsar wind nebulae, self-generated turbulence has been proposed as an explanation for the observed inhibited diffusion regions and the resulting $\gamma$-ray halos \cite{Evoli:2018aza,Fang:2019iym,Mukhopadhyay:2021dyh}. In the context of SNRs, it has been suggested that self-generated turbulence could affect the grammage \cite{DAngelo:2017rou,Recchia:2021vfw} or account for the flattening of CR spectra at low energies indicated by Voyager data \cite{Jacobs:2021qvh}, and in dwarf galaxies, self-generated turbulence may enhance radio signatures from dark matter annihilation \cite{Regis:2023rpm}.
In the specific context of Galactic CR propagation, self-generated turbulence is discussed as the explanation for the observed break in the spectra of CR nuclei at a few hundred GV \cite{Blasi:2012yr,Aloisio:2013tda,Aloisio:2015rsa,Evoli:2018nmb,Dundovic:2020sim}. Due to the numerical complexity involved, these models are typically only solved in one spatial dimension, resulting in limited precision when comparing them to CR data. Nevertheless, these models consistently suggest a characteristic feature: a continuous change in the slope of the diffusion coefficient, which is in contrast with the assumption of a sharply broken power law often employed in phenomenological models \cite{Genolini:2019ewc,Boschini:2020jty,Korsmeier:2021bkw}
Furthermore, these models also provide predictions for the spatial inhomogeneity of the diffusion coefficient.

In this study, we investigate CR propagation models, considering rigidity-dependent and spatially-dependent diffusion, as motivated by the works of Refs. \cite{Blasi:2012yr,Aloisio:2013tda,Aloisio:2015rsa,Evoli:2018nmb,Dundovic:2020sim}. 
We utilize data obtained from the AMS-02 experiment of primary and secondary nuclei ranging from protons to oxygen, complimented with additional measurements obtained at higher energies up to 5 TeV/nucleon from experiments such as CALET and DAMPE. We conduct global fits to determine the CR propagation parameters, taking into account smooth transitions in the slope of the diffusion coefficient.
Intriguingly, we observe a clear preference for models with a smooth break in the diffusion coefficient, which aligns closely with the characteristics of self-generated diffusion models as shown in Figure \ref{fig:main_result}. 

The paper is organized as follows:
In Sec.~\ref{sec:prod_prop} we summarize the model for production and propagation of CRs in the Galaxy. Then, in Sec.~\ref{sec:analysis} we detail the dataset and analysis technique we use to perform the fit to the propagation models. Finally, we present our results in Sec.~\ref{sec:results} before concluding in Sec.~\ref{sec:conclusions}.

\section{Cosmic-ray production and propagation}
\label{sec:prod_prop}

The propagation of CRs through the Galaxy involves complex interactions with various components, including magnetic field, photons, gas of the interstellar medium (ISM), energy losses, and Galactic winds. To model these interactions, a chain of coupled diffusion equations is used, one for each CR isotope. In this work, we employ the same model as in Ref.~\cite{Korsmeier:2021brc}, to which we refer for more comprehensive details. Here, we only summarize the key ingredients and highlight specific modifications implemented for this study. 

The main process governing the transport of Galactic CR nuclei is diffusion. To capture the energy dependence of the diffusion coefficient, we adopt a double-broken power law model as a function of rigidity, $R$:
\begin{eqnarray}
    \label{eqn:diffusion_coefficient}
     D(R) \propto \beta R^{\delta_l}
 	 \!\!&\cdot&\!\! \left( 1 + \left(\frac{R}{R_{D,0}}\right)^{\frac{1}{s_{D,0}}} \right)^{s_{D,0}\,( \delta - \delta_l) }  \\ \nonumber 
 	\!\!&\times&\!\! \left( 1 + \left(\frac{R}{R_{D,1}}\right)^{\frac{1}{s_{D,1}}} \right)^{s_{D,1}\,( \delta_h - \delta) },
\end{eqnarray}
where $\beta$ is the particle velocity in units of the speed of light, $R_{D,0}$ and $R_{D,1}$ are the two rigidity breaks, and  $\delta_l$, $\delta$, and $\delta_h$ are the power-law index below, between, and above the breaks, respectively. 
We allow smooth transitions at the positions of the breaks, which are controlled by the parameters $s_{D,0}$ and $s_{D,1}$. It is worth noting that these parameters enable considerable distortion and provide flexibility in shaping the diffusion coefficient, up to a point where, at first glance, it does not exhibit a clear power-law behavior anymore.
The normalization of the diffusion coefficient defined in Eq.~\eqref{eqn:diffusion_coefficient} is determined by the condition $D(R=4,\rm{GV})=D_0$. 

\begin{figure}[t]
\centering
\includegraphics[width=0.49\textwidth]{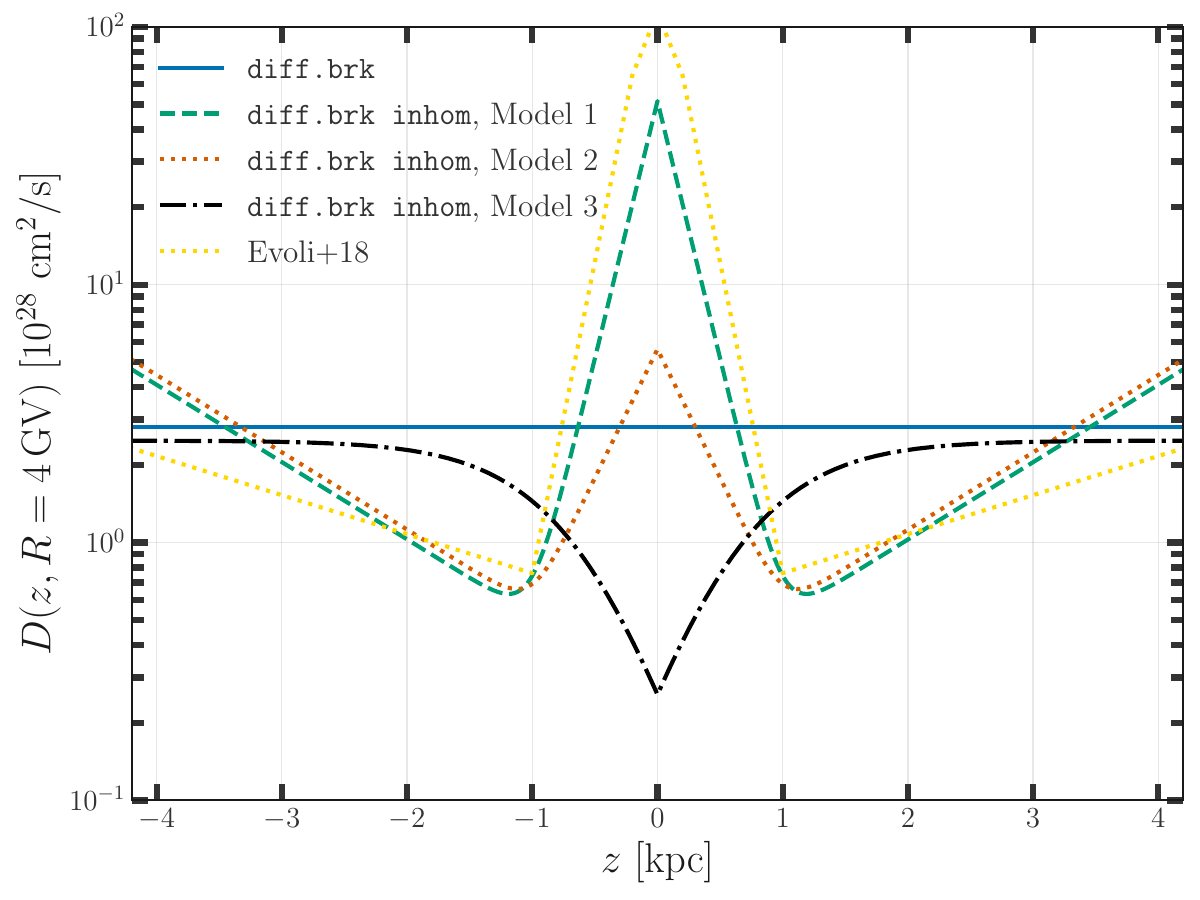}
\caption{Diffusion coefficient as a function of distance to the galactic plane, $z$, considered for the three benchmark cases of the \dbin\ framework. The diffusion coefficient is evaluated at $R=4$ GV and for the best-fit values found in Sec.~\ref{sec:results}. We also show, as a yellow dotted curve, the prediction from \cite{Evoli:2018nmb} based on a model of self-generated turbulence.}
\label{fig:Dz}
\end{figure}

The most common approach, which is also taken in Ref.~\cite{Korsmeier:2021brc}, assumes a homogeneous diffusion coefficient within the diffusion halo. However, in this study, we move beyond this simplification and introduce a novel aspect by allowing the normalization of the diffusion coefficient to vary as a function of the distance from the Galactic plane, denoted as $z$:
\begin{eqnarray}
\label{eqn:diffusionz}
     && \log_{10} \big( D_{0,z}(z)/D_0 \big) =  z_{m,1} \cdot (|z|-z_{br}) \\ \nonumber
     && \qquad + z_{s} (z_{m,2}-z_{m,1}) \log{ \left( 1+\left(\frac{\exp{(|z|)}}{\exp{(z_{br})}}\right)^{1/z_s}\right) } .
\end{eqnarray}
Here $z_{m,1}$ and $z_{m,2}$ are the slopes of $\log_{10} \big( D_{0,z}(z) \big)$ below and above a transition at $z_{br}$, while the parameter $z_s$ enables a smooth transition.
The parametrization we employ offers sufficient flexibility to capture the shape of the diffusion coefficient as derived in the model of self-generated turbulence from Ref.~\cite{Evoli:2018nmb}. 
To explore and evaluate various scenarios, we consider different benchmarks for the functional form of $D_{0,z}(z)$, which we shown in Figure \ref{fig:Dz} normalized to the best-fit found in Sec.~\ref{sec:results}.
The value of the diffusion coefficient is forced to vary in a range between $10^{26}$ and $10^{31}$ cm$^2$/s. 
For convenience, in the following we will show the results for the rigidity dependence of $D$ at $z=3$ kpc, where, as seen from Figure \ref{fig:Dz}, the different models have approximately the same $D(z)$ value. The choice $z=$ 3 kpc is representative because CRs spent most of their time away from the Galactic plane.
We show in Fig.~\ref{fig:Dz} also the prediction for $D(z)$ as taken from Ref.~\cite{Evoli:2018nmb}, which is qualitatively similar to the one we use for Model 1 and 2 (see Tab~\ref{tab:Dz} for the model parameters).
We have verified that the shape of $D(z)$  taken from Ref.~\cite{Evoli:2018nmb} is very similar to our model 1 when considering rigidity values ranging from $\sim$ GV to 1000 GV.
Above $\sim$ 1 TV, instead, the model from Ref.~\cite{Evoli:2018nmb} has a slight transition and becomes more similar to our model 2. This feature of an energy-dependent $D(z)$ shape might become interesting and important in the future when more precise B/C data above $\sim$ TV will become available.

In addition to diffusion, several other physical processes are incorporated into the propagation equation, including convection which is modeled by a velocity vector orthogonal to the Galactic plane $\bm{V}(\bm{x})= {\rm sign}(z) v_{{\rm c}}\,{\bm e}_z$, reacceleration which is characterized by momentum-space diffusion through the parameter $D_{pp} \sim v_\mathrm{A}^2/D_{xx}$, where $v_\mathrm{A}$ represents the speed of Alfv\'en magnetic waves, and energy losses in the ISM. 
The source term for each primary cosmic ray species $i$, originating from astrophysical sources, is determined by $q_i(\bm{x}, p) = Q_i(R) \rho(\bm{x})$, where $Q_i(R)$ represents the rigidity-dependent term corresponding to the energy spectrum at injection for each species and  $\rho(\bm{x})$ the source term density. To parameterize $Q_i(R)$, we adopt a smoothly broken power-law:
\begin{equation}
 \label{eqn:energy_spectrum}
 Q(R) = Q_{0,i} R^{\gamma_{1,i}} \left( 1 + \left(\frac{R}{R_\mathrm{inj}}\right)^{1/s_{\rm{inj}}} \right)^{s_{\rm{inj}}\,( \gamma_{2,i} - \gamma_{1,i}) }  ,
\end{equation}
where $R_\mathrm{inj}$ is the break rigidity, and $\gamma_{1,i}$ and $\gamma_{2,i}$ are the two spectral indices above and below the break. The smoothing of the break is parameterized by $s_{\rm{inj}}$. For the spatial distribution of sources  $\rho(\bm{x})$ we assume the one of SNRs reported in Ref.~\cite{Green:2015isa}.

Finally, we include solar modulation, using the force-field approximation \cite{Fisk:1976aw}, which is fully determined by the solar modulation potential $\varphi$.
We use the same $\varphi$ for all the CRs except for antiprotons for which we use a different value to account for the evidence of a charge dependence in the solar modulation (see, e.g., \cite{Cholis:2020tpi,PhysRevLett.130.211001,AMS:2018avs,AMS:2023hom}).

We investigate two frameworks for CR propagation:
\begin{itemize}
\item \db: 
The diffusion coefficient in our model is represented by a double-broken power law, as shown in Eq.~\ref{eqn:diffusion_coefficient}. The first break typically occurs within the range of 1--10 GV, while the second break, directly observed in cosmic ray spectra, appears at approximately 200--400 GV \cite{Weinrich:2020cmw,Korsmeier:2021bkw,Korsmeier:2021brc,Vecchi:2022mpj}.
This setup incorporates convection with a fixed velocity $v_c$, while reacceleration is assumed to be negligible and discarded. The CR injection spectra are taken as single power laws ($\gamma_1=\gamma_2$ in Eq.~\eqref{eqn:energy_spectrum}), where the spectral indices for protons ($\gamma_p$), helium ($\gamma_{\rm{He}}$), and CNO nuclei ($\gamma_{\rm{CNO}}$) are adjusted individually. This individual adjustment is necessary because assuming the same power law for all species significantly worsened the agreement with the data \cite{Korsmeier:2021bkw,Korsmeier:2021brc,Genolini:2019ewc,Genolini:2021doh}.

As a default, we consider homogeneous diffusion, namely $D_{0,z}(z)=D_0$.
Thus, the free parameters in this setup are: the spectral indices $\gamma_p$, $\gamma_{\rm{He}}-\gamma_p$, and $\gamma_{\rm{CNO}}-\gamma_p$ for the injection spectrum; $D_0$, $\delta_l$, $\delta$, $\delta_h$, $R_{D,0}$, $R_{D,1}$, $s_{D,0}$, and $s_{D,1}$ for the diffusion coefficient; $v_{0,c}$ for convection; and a single solar modulation potential $\phi$ that applies to all cosmic ray species, except $\bar{p}$ which has a solar modulation potential $\bar{\phi}$.

\item \ib: 
In contrast to the previous setup, this model adopts a smoothly broken power-law for the injection spectrum with a break at a few GV, allowing for individually free spectral indices for protons, helium, and CNO nuclei above and below the break. The rigidity break and smoothing are assumed to be the same for all species. Moreover, in this framework, we employ only a single broken power law with a smooth transition for the diffusion coefficient, with the break at a few hundred GV. Finally, reacceleration is included via the Alfv\'en velocity parameter. Convection is turned off.

Thus the free parameters are $\gamma_{1,p}$, $\gamma_{2,p}$, $\gamma_{1,\rm{He}}$, $\gamma_{2,\rm{He}}$, $\gamma_{1,\rm{CNO}}$ and $\gamma_{2,\rm{CNO}}$ for the spectral indices of proton, helium and CNO injection, $R_{\rm{inj}}$ and $s_{\rm{inj}}$ for the rigidity break and smoothing in the injection spectrum; $\delta$, $\delta_h$, $R_{D,1}$, and $s_{D,1}$ for the parameters related to diffusion; and $v_{\rm{A}}$ for the strength of reacceleration.
\end{itemize}

In the \db\ framework, we explore the possibility of inhomogeneous diffusion and refer to it as \dbin. Specifically, we investigate three benchmark scenarios in addition to the homogeneous case.
We show in Tab.~\ref{tab:Dz} the values for the parameters of the function $D_{0,z}(z)$ (see Eq.~\ref{eqn:diffusionz}).

The first two scenarios are inspired by the model of self-generated diffusion from Ref.~\cite{Evoli:2018nmb}. In these cases, the diffusion coefficient is significantly larger near the Galactic plane and reaches a minimum value around $z=1$~kpc. The difference between the first two models lies in the degree to which the diffusion coefficient increases at $z=0$. This increase can be explained if the advection of the magnetic turbulence injected by SNRs happens faster than the cascading of the turbulence to smaller scales.

Conversely, the third benchmark scenario exhibits a behavior opposite to the previous two. Here, the diffusion coefficient is suppressed near the Galactic plane and then gradually increases and saturates for $z\gtrsim 1.5$ kpc. This behavior is motivated by observations of $\gamma$-ray halos surrounding the brightest Galactic pulsar wind nebulae \cite{HAWC:2017kbo,DiMauro:2019yvh,DiMauro:2019hwn,DiMauro:2020jbz}. The extension of these halos are compatible with $\gamma$ rays produced through inverse Compton scattering of electrons and positrons emitted by the pulsar with Galactic photons in an environment with inhibited, i.e.,~small, diffusion. 
Considering that the Galaxy contains millions of pulsars, the diffusion coefficient in the Galactic disk could be orders of magnitude smaller than the one in the halo. Such a scenario has been recently explored in Ref.~\cite{Liu:2018ujp,Johannesson:2019jlk,Zhao:2021yzf,Jacobs:2023zch}.
In particular, the distribution of Galactic pulsars and supernova remnants goes as approximately $\exp(-z/z_0)$ (see, e.g., \cite{Lorimer:2006qs}) where $z_0$ is between 0.1 and 0.5 kpc. 
Therefore, we expect  the diffusion coefficient
to follow a similar trend, being suppressed in the disc and then approaching the average one in the Galactic halo with a transition scale of about $z_0$.
This motivates our choice for the shape of $D(z)$ for Model 3.

\begin{table}
\begin{center}
\begin{tabular}{cc cc cc cc c}
\hline
\hline
Model  &$\quad$& $z_{m,1}$ &$\ $& $z_{m,2}$ &$\ $& $z_{br}$  &$\ $&  $z_s$ \\
\hline
 1 && -2.0 && 0.3 && 1.0 && 0.1  \\
 2 && -1.0 && 0.3 && 1.0 && 0.1  \\
 3 &&  1.5 && 0.0 && 0.5 && 0.5  \\
\hline
\end{tabular}
\caption{This table shows the value of the parameters $z_{m,1}$, $z_{m,2}$, $z_{br}$ and $z_s$ of the function $D_{0,z}(z)$ that parametrize the variation of the diffusion coefficient with the $z$ direction.}
\label{tab:Dz}
\end{center}
\end{table}

There has been increasing awareness in recent years of the importance of the uncertainties in the nuclear cross sections for the production of secondary CRs \cite{Korsmeier:2016kha,Genolini:2018ekk,Korsmeier:2021bkw,Genolini:2023kcj,DiMauro:2023oqx}.
To account for these unknowns, when fitting the model to the AMS-02 data, besides the parameters related to the physics of CR propagation, we include further parameters for the uncertainties in the nuclear fragmentation cross sections. The latter are treated as nuisance parameters. To parameterize the cross-section uncertainties governing the production of secondary CRs, we introduce a re-normalization factor and a change of slope of the cross-section below a fixed kinetic energy per nucleon of 5 GeV/n. For, e.g., Be
these two parameters would read as
as $A_{\rm{XS}} \rightarrow$ Be, and  $\delta_{\rm{XS}} \rightarrow$ Be. Analogous parameters are introduced for all the secondary production cross sections. 
Regarding antiprotons we adopt the default cross-section parameterization given in \cite{Korsmeier:2018gcy}, and we only introduce as a nuisance the overall renormalization, $A_{\bar{p}}$.
A summary of the fit parameters and their corresponding priors for each model is provided in Appendix \ref{sec:app}. We assume flat priors for all the parameters.

\section{Methodological Framework}
\label{sec:analysis}

The framework employed in this paper is the same as the one outlined in previous works \cite{Korsmeier:2018gcy,Korsmeier:2021bkw,Korsmeier:2021brc,DiMauro:2023oqx}. In the following, we summarize the key points of this framework.

We use the most recent version, v.~57, of the \textsc{Galprop} code\footnote{http://galprop.stanford.edu/} \cite{Strong:1999sv,2009arXiv0907.0559S,Porter:2021tlr} for numerical solutions of the CR propagation equations. Our study specifically assumes cylindrical symmetry and, thus, incorporates two spatial dimensions:  distance from the Galactic plane, $z$, and Galactocentric radius, $r$, up to  $r_{max}=$ 20 kpc. The half-height of the diffusion halo, denoted as $L$, is set to a fixed value of 4.2 kpc. 
We have, nonetheless, checked that our results for the shape in rigidity of the diffusion coefficient $\delta(R)$ as well as the value of most of the propagation and cross section parameters are essentially independent of the exact choice of $L$. This is partly due to the well-known degeneracy between the $L$ and the normalization of the diffusion coefficient \cite{Evoli:2019iih,DiMauro:2023oqx}, which nominally applies to homogenous diffusion, but holds also for inhomogeneous diffusion at the current level of precision of the data.
We have also checked that using larger values of $L$ the best-fit $\chi^2$ value of the fits remains basically unchanged, indicating that the data are currently unable to constrain well $L$, in agreement with previous works \cite{Korsmeier:2021brc}.
Solving the diffusion equations we include CR nuclei up to silicon.

We fit the absolute fluxes of $p$, He, C, O, and N, as well as the ratios B/C, Be/C, Li/C, and $\bar{p}/p$, using data obtained by AMS-02 after 7 years of data taking \cite{AGUILAR2020}. Additionally, we incorporate AMS-02 data for the $^3$He/$^4$He ratio \cite{2019PhRvL.123r1102A}. The ratios of secondaries to primaries, such as $\bar{p}/p$, $^3$He/$^4$He, Li/C, Be/C, and B/C, play a crucial role in determining the propagation parameters, while the data for $p$, He, C, N, and O determine the injection spectra.

To expand our data set at higher energies, we include the $p$ data obtained by CALET \cite{PhysRevLett.129.101102} and the He data from DAMPE \cite{Alemanno:2021gpb}. 
However, given the potential break observed by the above two experiments at about 10 TeV,
we restrict the data to energies below 10 TeV in rigidity to avoid introducing unnecessary complexity that lies beyond the scope of our interest.
Furthermore, we expand the B/C data measured by AMS-02 by incorporating recently published DAMPE data \cite{DAMPECOLLABORATION20222162}, which covers kinetic energies per nucleon up to 5 TeV/nuc. Considering that the B/C data from DAMPE and AMS-02 may be subject to different systematic effects, such as variations in the calibration of the energy scale, we introduce a renormalization factor for the DAMPE data, allowing for a relative adjustment of the normalization with a prior range of $0.90-1.10$. 

Finally, we incorporate $p$ and He data from Voyager \cite{stonevoyager2013} above 0.1 GeV/nuc to calibrate the interstellar injection spectrum. Voyager data is utilized only above 0.1 GeV/nuc to avoid complications arising from very low energies, such as stochasticity effects resulting from local sources or the potential presence of an additional low-energy spectral break \cite{Phan:2021iht}.

The spectra of AMS-02 are affected  by solar modulation. 
Since all measurements correspond to the same data-taking period we use a single 
Fisk potential for all the species, except the $\bar{p}/p$ ratio to account for the effect of charge-sign dependent Solar modulation.
The data from Voyager are taken outside the Heliosphere and therefore not affected by solar modulation while for DAMPE and CALET data, which are at very high energies, solar modulation is negligible. 

The analysis incorporates a total of 691 data points. With the model having between 25 and 30 free parameters, a good-fit $\chi^2$ value is expected to be on the order of or below 650.

To sample the parameters of the CR propagation model, we utilize the nested sampling algorithm implemented in \textsc{MultiNest} package~\cite{Feroz:2008xx}. This approach allows us to obtain posterior distributions as well as the Bayesian evidence. Our settings for the nested sampling algorithm are 400 live points, an enlargement factor (\textsc{efr}) set to 0.7, and a stopping criterion (\textsc{tol}) of 0.1.

\section{Results}
\label{sec:results}

In this section, we first provide an overview of the results derived from the fitting of the CR data using two distinct frameworks, \db\ and \ib\ as detailed in Sec.~\ref{sec:prod_prop}, before we turn to our central findings, i.e., the results regarding the  shape of the diffusion coefficient as a function of rigidity as well as its spatial dependence.

\subsection{Overview of the results}
\label{sec:CRfit}

\begin{figure*}[t]
\centering
\includegraphics[width=0.33\textwidth]{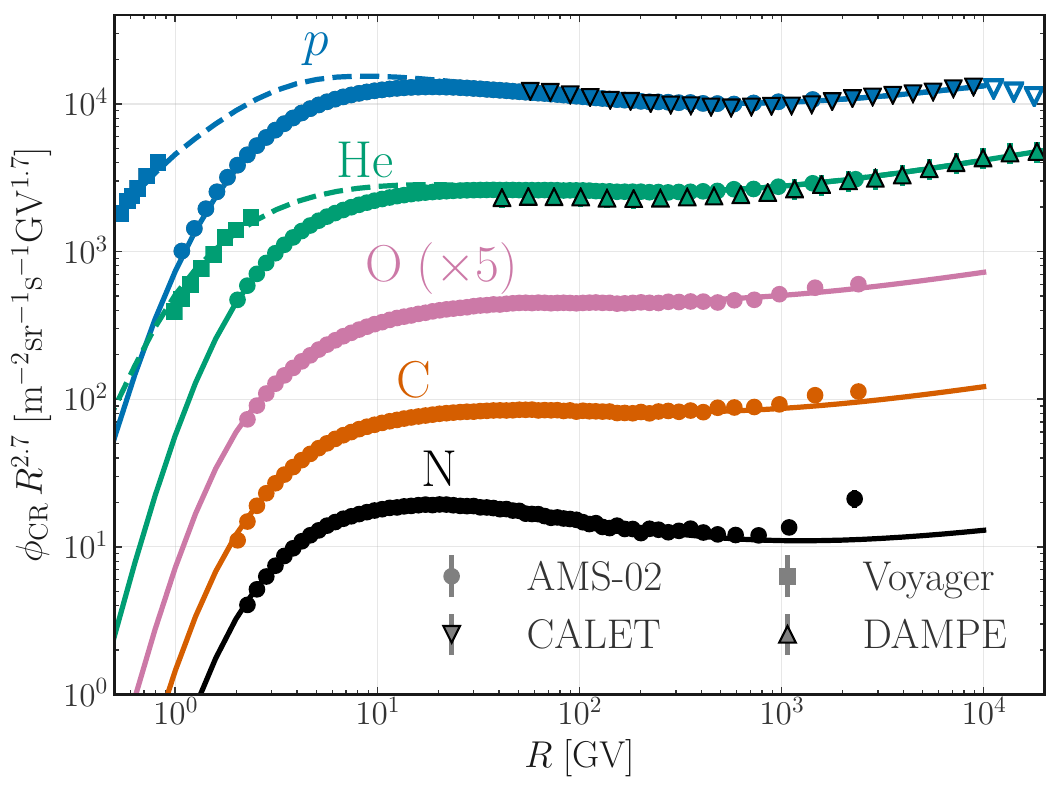}%
\includegraphics[width=0.33\textwidth]{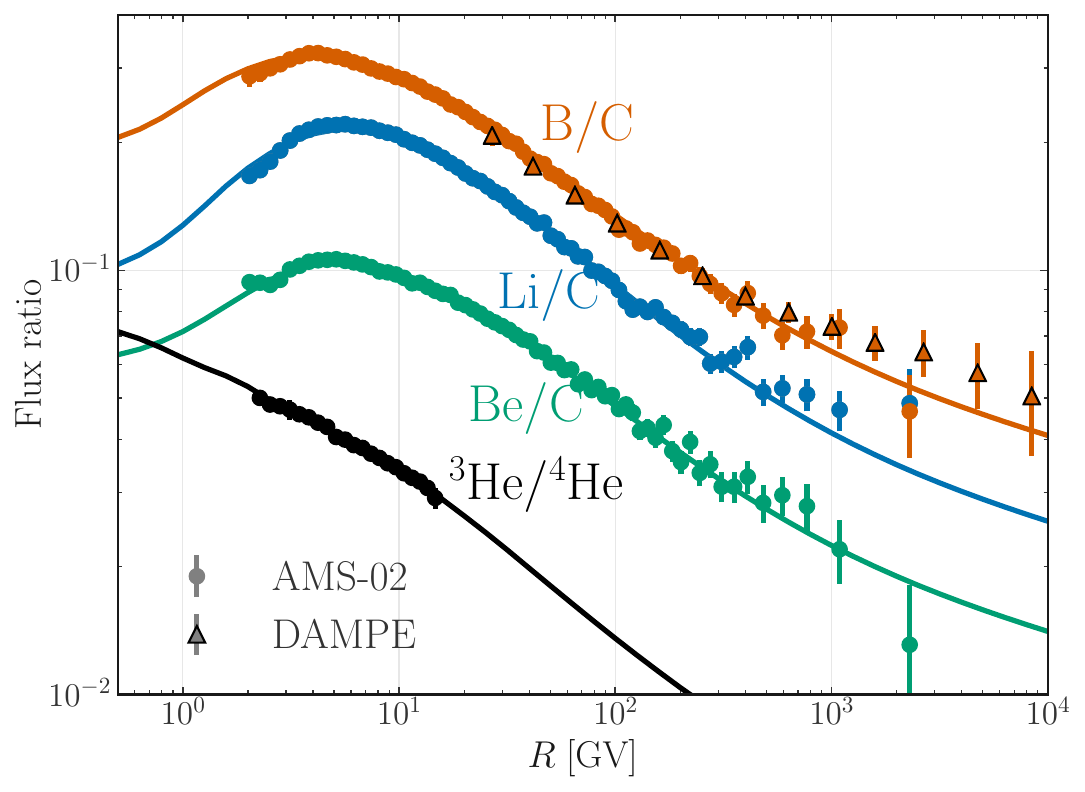}%
\includegraphics[width=0.33\textwidth]{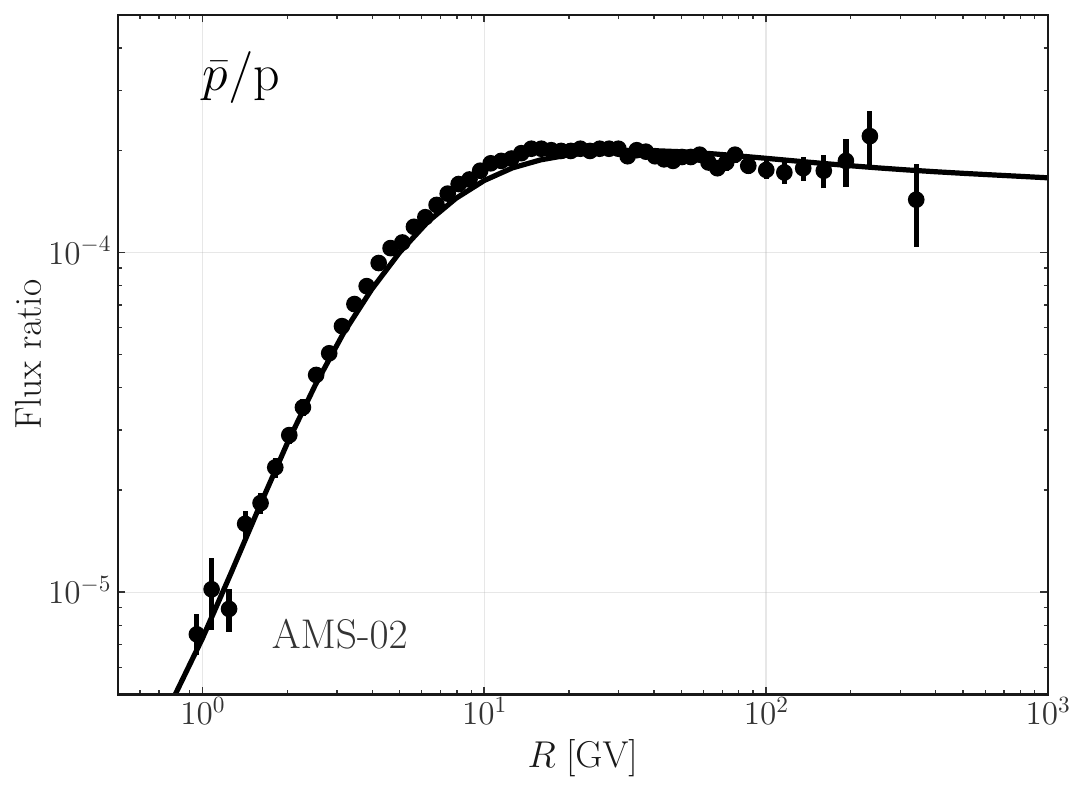}
\caption{
   Comparison of the CR fluxes and flux ratios of the \db\ model to the fitted data sets from the AMS-02, CALET, DAMPE, and Voyager experiments.}
  \label{fig:CRfit}
\end{figure*}

Figure \ref{fig:CRfit} shows the comparison between the data and the best-fit results obtained for the default \db\ model, which assumes homogeneous diffusion throughout the Galaxy. The model effectively captures the essential characteristics of the various datasets, offering a good overall description. The total $\chi^2$ of the fit is equal to 404, resulting in a reduced chi-square value of approximately $\tilde{\chi}^2=0.6$ ($\tilde{\chi}^2=\chi^2/d.o.f.$, where $d.o.f.$ represents the number of data points minus the free parameters of the model). 
The $d.o.f.$ for all fits and models considered in this work are about 650, with the exact values given in the appendix. 
A well-fitting model would typically yield $\tilde{\chi}^2\approx1$. The observed deviation from this expectation is probably related to correlations in the systematic uncertainties present in the AMS-02 CR data. Although the AMS-02 collaboration does not provide explicit details regarding these correlations, when modeled and estimated, their inclusion typically increases the $\chi^2$ value and simultaneously reduces the uncertainty associated with the propagation parameters. In this sense, neglecting the correlations is a conservative approach. For a more comprehensive discussion on this aspect, we refer to Refs.~\cite{Boudaud:2019efq,Heisig:2020nse,Cuoco:2019kuu,DiMauro:2023oqx}.

When looking at the results for the individual data sets, shown in the table in the Appendix, indeed, it is apparent that only the AMS-02 spectra have low $\tilde{\chi}^2$ values.
For example, CALET and DAMPE data for $p$ and He respectively have $\tilde{\chi}^2$ values close to 1, while DAMPE data for B/C yields a $\tilde{\chi}^2$ of 1.5.
The latter slightly high value is due to the fact that the fit does not completely account for the hardening observed by DAMPE in the B/C spectra at high energies, as shown in Fig~\ref{fig:CRfit}.  
The deviation, however, is only very mild.
The fit to the Voyager data for $p$ and He results in slightly larger values of $\tilde{\chi}^2$, ranging between 2 and 3. However, we note that the Voyager data cover energy ranges well below the primary focus of this study. Achieving a better fit to the Voyager data typically requires an additional low-energy break in the injection spectra, as discussed in the Refs.~\cite{Vittino:2019yme, Phan:2021iht}.

In agreement with the results presented in Ref.~\cite{DiMauro:2023oqx}, a small level of convection is required, with the best-fit value for the convection velocity of $\sim$13 km/s. The injection spectrum of $p$ exhibits a slope of 2.37, while He and the CNO group exhibit slightly, but significantly, harder spectra with indices of 2.31 and 2.35, respectively.
The Fisk potential, which parameterizes the solar modulation, takes a value of 0.46 GV for all the CRs except for $\bar{p}$ for which the best fit is 0.74 GV. This result confirms the presence of a charge-sign dependence of solar modulation.

The nuisance parameters for the fragmentation cross section converge to values close to the default parameterization. However, there are deviations for the renormalization of Li, which takes a value close to  the upper edge of the prior distribution at 1.30, and for C, which tends to prefer smaller values around 0.77. These deviations from the defaults have been previously documented and discussed in the literature \cite{Maurin:2022irz,Boschini:2019gow,DiMauro:2023oqx}.
The normalization of the production cross section for antiprotons ($\bar{p}$) has a best-fit value of 1.13, which is also consistent with the current theoretical uncertainties \cite{diMauro:2014zea,winkler_2017,Korsmeier:2018gcy}.

\bigskip

The $\chi^2$ of the \ib\ model is slightly better than the one of the \db\ model. The improvement is driven by the data of $p$, He, C, N, and O, which give reduced $\chi^2$ values of approximately 0.25 to 0.30. This result can be explained by the fact that this model has more freedom in the shape of the primary CR injection spectra. 
Conversely, for the secondary-to-primary data, the overall goodness of fit is comparable to that of the \db\ model. However, the \ib\ model captures the hardening of the boron-to-carbon (B/C) ratio observed above 1 TeV by DAMPE slightly better than the \db\ model in Ref.~\cite{Ma:2022iji}.

The \ib\ model requires a significant amount of reacceleration, as seen from the best-fit value for the Alfv\'en velocity of $\sim$25 km/s. 
The injection spectrum above 20 GeV closely resembles that of the \db\ model. However, at lower energies, the \ib\ model converges to injection slopes around 1.8 and 2.0 with a smooth break occurring at 8--9 GeV and a smoothing parameter which is 0.3.
The Fisk potential for all the CRs exhibits a value of 0.66 GV while for $\bar{p}$ it is 0.50, consistent with Ref. \cite{Korsmeier:2021bkw}.

The \ib\ model requires substantial deviations of the fragmentation cross sections from the default parametrization. 
The renormalization of Li goes to the upper edge of the prior, as for the \db\ model, while the slope parameters $\delta_{\rm XS}$ have positive tilts of about 0.1--0.3.
Some of the preferred nuisance parameters push the fragmentation cross section into a regime that exceeds the uncertainties allowed by the data, making this model less favored overall, an observation that aligns with Ref.~\cite{Korsmeier:2021bkw,DiMauro:2023oqx}.
In particular, the model pushes for a very low value of the secondary Carbon production cross section of about 0.3.

\bigskip

\begin{figure}[t]
  \centering
  \includegraphics[width=0.49\textwidth]{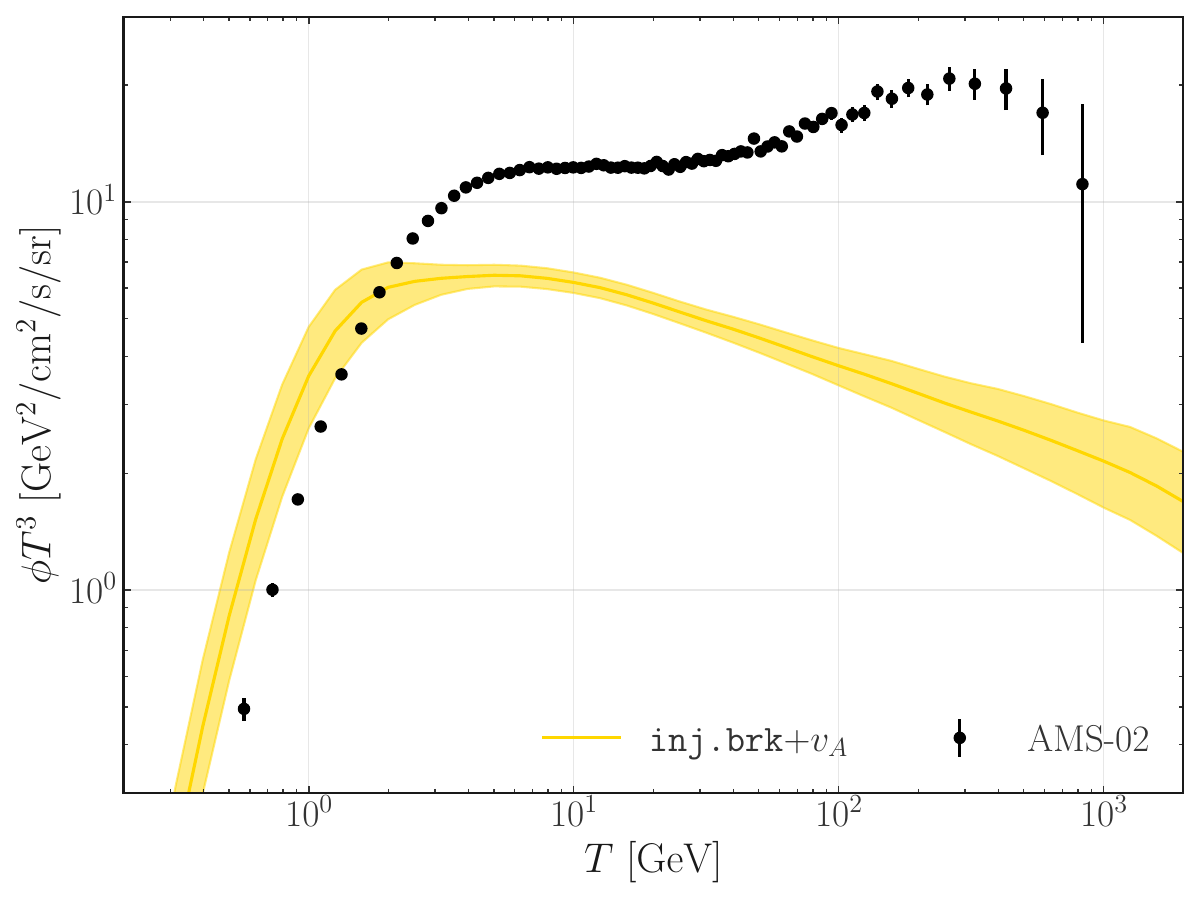}
  \caption{Prediction of the secondary positron flux as a function of the kinetic energy $T$ for the model \ib\ compared to the AMS-02 data. The band  includes the uncertainty related to propagation parameters and the production cross section of $e^+$.}
  \label{fig:pos}
\end{figure}

Positron data at low energy can be used to constrain propagation models and the size of the diffusive halo as previously noted for example in Refs.~\cite{Weinrich:2020ftb,DiMauro:2023oqx}. Therefore, as a further test, we also calculate the prediction for the secondary positron flux and compare it with the low-energy data from AMS-02.
To account for the solar modulation, we employ the best fit of the Fisk potential obtained for primary CRs and checked that conclusions are not affected by changing to the one from $\bar{p}$.
For the \db\ or \dbin\ models we find very similar $e^+$ predictions as in \cite{DiMauro:2023oqx} and none of these models are in contradiction with AMS-02 $e^+$ data.
Instead, the theoretical prediction for the \ib\ case is a factor of about two above the data for energies below 2 GeV, as can be seen from Fig.~\ref{fig:pos}, indicating a significant tension.
Therefore, in conclusion, the reacceleration model seems to be disfavored by multiple tensions, both in the parameters for the nuclear cross sections needed to fit secondary CR data and by a too-large flux of secondary $e^+$ at energies below 10 GeV.

\begin{figure}
  \centering
\includegraphics[width=0.46\textwidth]{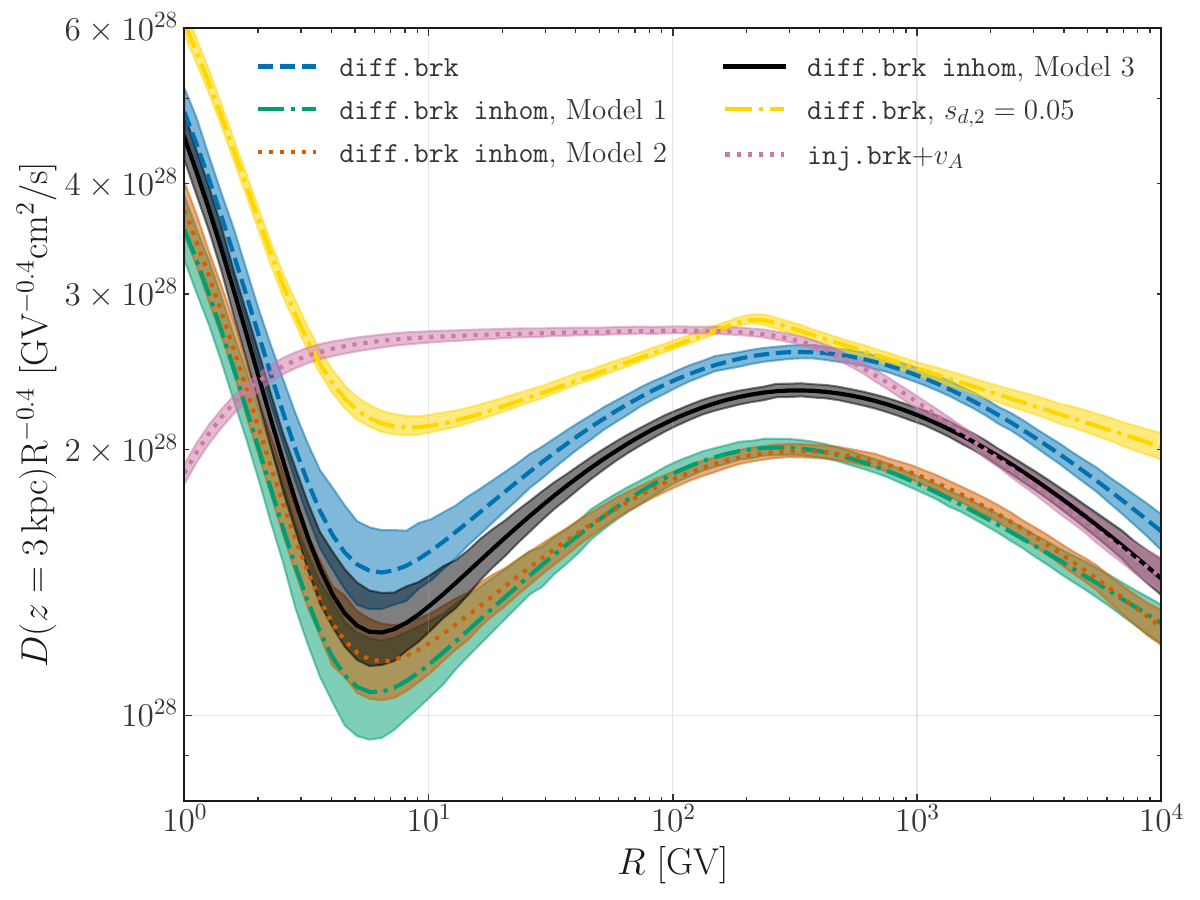}%
  \caption{
    Comparison of the diffusion coefficient evaluated at $z=3$ kpc for the different CR propagation models.}
  \label{fig:D0}
\end{figure}

\subsection{Results for the diffusion coefficient and smoothing parameters}
\label{sec:resultsD}

The main result of this analysis is the shape of the diffusion coefficient as a function of rigidity and Galactic $z$, in comparison to the predictions of models with self-generated turbulence. 
We find that in the \db\ and \dbin\ models, the diffusion coefficient exhibits a significant smoothing of both the low-energy and high-energy break. The low-energy break presents a smoothing of about $s_{d,0} = $ 0.25--0.30 while the high-energy break  a smoothing of about $s_{d,1}= $ 0.80--1.35. This results in a gradual change in the effective slope $\delta$ of the diffusion coefficient from a few GV to a few TV, as depicted in Figure \ref{fig:main_result} for the \dbin\ Model 2. The effective slope gradually decreases from 0.6 at a few GV to about 0.2 at a few TV. The behavior qualitatively resembles the expectations of self-generated turbulence models~\cite{Blasi:2012yr,Aloisio:2013tda,Aloisio:2015rsa,Evoli:2018nmb,Dundovic:2020sim}.
Instead, the cases without smoothing ($s_{d,0}=s_{d,1}=0.05$) present a very sharp transition between the $\delta$ value below and above the rigidity breaks which are inconsistent with the shape of $\delta(R)$ obtained with self-generated turbulence models.
The propagation setup with reacceleration \ib\, instead, assumes only a single break in diffusion at high energy. The smoothing for this break of $s_{d,1}=0.34 \pm 0.08$ is less pronounced than in the \db\ case. The low rigidity break is instead replaced with a break in the CR injection spectrum, which features a smoothing of $s_{inj}=0.33 \pm 0.03$.

\begin{figure*}[t]
  \centering
  \includegraphics[width=0.49\textwidth]{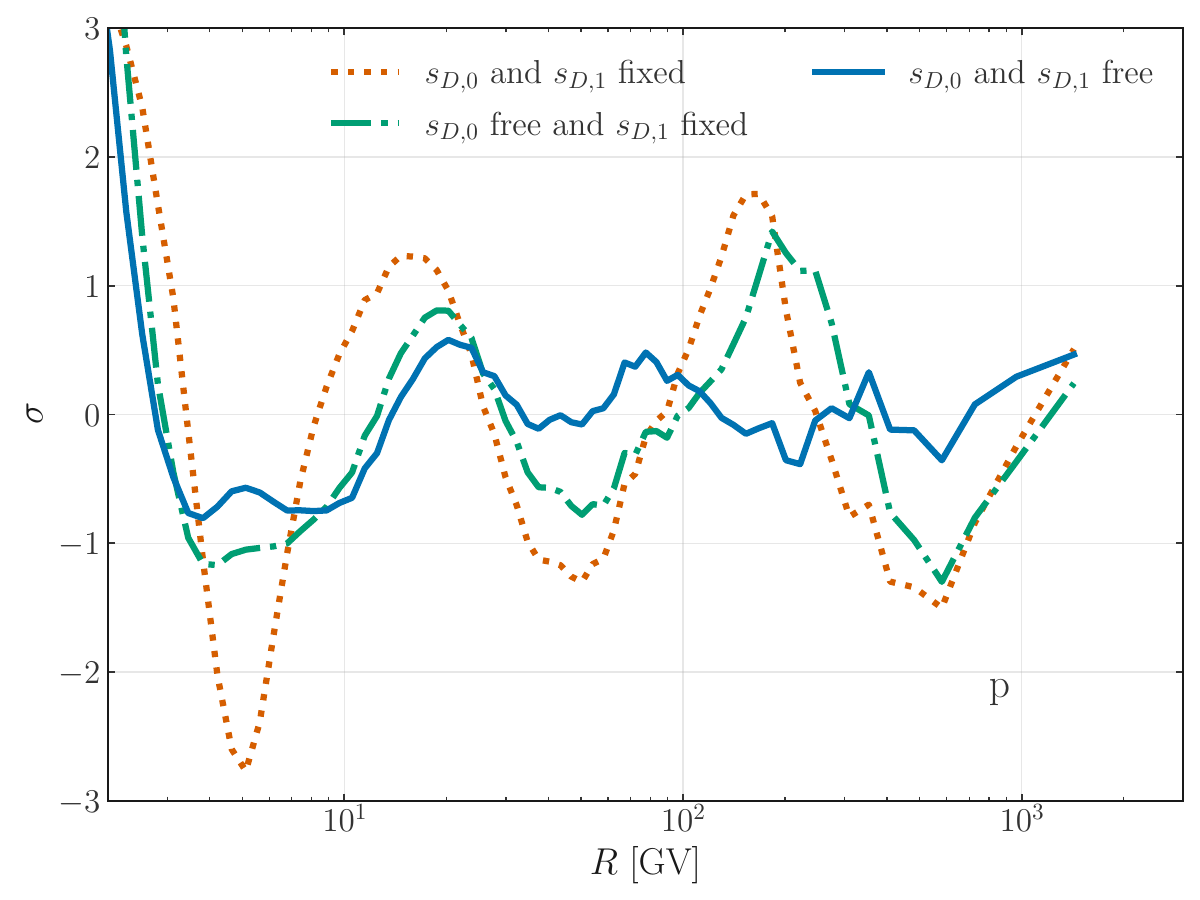}
  \includegraphics[width=0.49\textwidth]{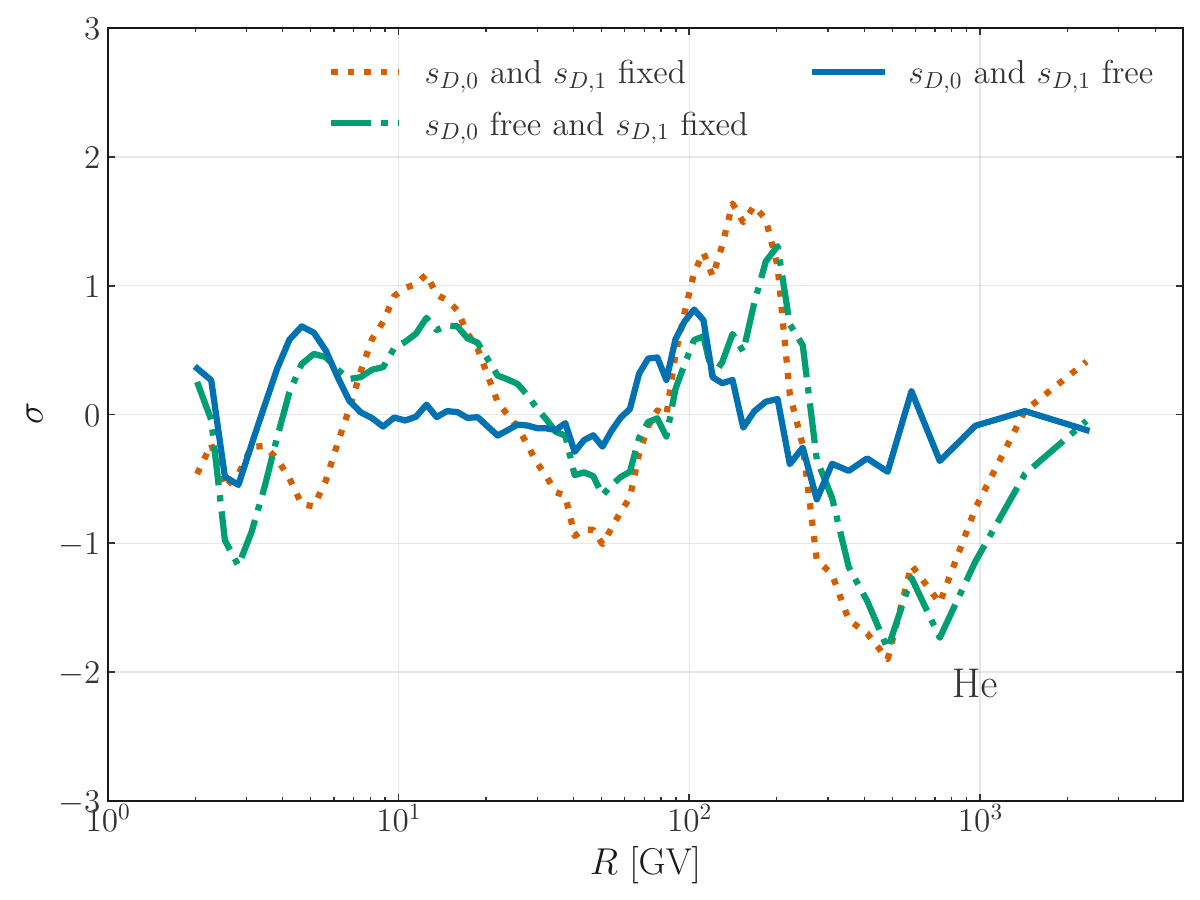}
  \caption{
    Residuals in units of $\sigma$ = (data-model)/error for the \db\ model for AMS-02 the proton flux data (left panel) and helium flux (right panel). The results are shown for three cases:
    First both smoothing parameters, $s_{d,0}$ and $s_{d,1}$, are fixed to a sharp break (orange dotted lines), then smoothing is introduced for the low-energy break keeping a sharp transition at high energy  (green dashed-dotted lines), and finally smoothing in both transitions is considered blue solid lines).
  }
  \label{fig:CRrespHe}
\end{figure*}

In Fig.~\ref{fig:D0}, we show the shape of the diffusion coefficient as a function of rigidity for the different models explored in this study.
For the inhomogeneous models, we show $D$ evaluated at $z=3$ kpc, i.e. $D(R,z=3\,\rm{kpc})$, because CRs spend most of the time propagating away from the disk and at $3$ kpc the different models for $D(z)$ provide a similar value as can be seen in Fig.~\ref{fig:Dz}.
It is evident that the shape obtained when fixing $s_{d,0}$ and $s_{d,1}$ to 0.05 differs significantly from the shapes obtained when allowing these parameters to vary freely. The model with fixed smoothing parameters has also an overall larger $D$ with respect to the other cases. In contrast, the cases tested with homogeneous or inhomogeneous diffusion in the $z$ direction exhibit similar smooth changes in slope with rigidity.
We note that the relative normalization of $D(R)$ obtained for various \db\ and \dbin\ models depends on the choice of the reference $z$ (see Fig.~\ref{fig:Dz}).

To assess the significance of a smooth change of $\delta$ compared to a sharply broken power law, for the \db\ model we perform additional fits where the low-energy and/or high-energy break are fixed to a sharp transition. In the \ib\ framework, we do the same but for the low-energy break in injection and the high-energy break in diffusion. The  fit results, as well as the best-fit values for the indices of the diffusion coefficient and the smoothing parameters, are presented in Tab.~\ref{tab:fitresultsD}.
When both smoothing parameters $s_{d,0}$ and $s_{d,1}$ are fixed for the \db\ model, a $\chi^2$ value of 582 is obtained and the slopes  $\delta_l, \delta$ and $\delta_h$ are $-0.42$, 0.58, and 0.35 respectively, consistent with findings of Refs.~\cite{Maurin:2022irz}. However, allowing a smoothing in the low-energy break, as done in previous works \cite{Korsmeier:2021bkw,Korsmeier:2021brc,DiMauro:2023oqx}, leads to a significant improvement and a $\chi^2$ of 458. Finally, allowing a smooth transition also for the second break further improves the $\chi^2$ to 385. The smoothing of the second break alone improves the $\chi^2$ by 73, which, under the assumption that the Chernoff theorem applies to our analysis, corresponds to a significance of about $8.3\sigma$ for the presence of smoothing in the high-energy break\footnote{Based on the asymptotic theorem of Chernoff \cite{10.1214/aoms/1177728725}, the $\Delta \chi^2$ follows a $\chi^2$ distribution, from which the significance of the signal can be calculated.}.
The significance of the smoothing in the low-energy break is even higher, despite the amount of the smoothing being smaller. This conclusion is also supported by considering the Bayesian evidence $Z$ for the three models under consideration (see Tab.~\ref{sec:prod_prop}). Specifically, the natural logarithm of the ratio of $Z$ for the case with and without smoothing is about 41 indicating very strong evidence in favor of the presence of a smooth transition in the high-energy part of $D$ \cite{Trotta:2008qt}. The parameter $s_{d,0}$ converges to a value of $0.30\pm0.05$, while $s_{d,1}$ is found to be $1.43^{+0.20}_{-0.53}$. Consequently, the transition in the slope of the diffusion coefficient exhibits a considerably smoother behavior for the high-energy break compared to the low-energy break.
We point out that this is the first time an analysis has been performed to find the smoothing of the high-energy break in the diffusion coefficient. Previous papers have either chosen a sharp transition of the break \cite{Korsmeier:2021bkw,Korsmeier:2021brc} or run the fit for a small range for the prior on $s_{d,1}$ \cite{Maurin:2022irz,DiMauro:2023oqx}.

The improvement in the fit to the CR data is mainly driven by the fit to primaries $p$, He and O. This effect is illustrated in Fig.~\ref{fig:CRrespHe}, where we compare the residuals of the AMS-02 $p$ and He data for the three cases discussed earlier. The figure clearly shows that sharp transitions (i.e., the case with $s_{d,0}$ and $s_{d,1}$ fixed to 0.05) provide systematic features in the residuals, with  strong oscillations from negative to positive in the vicinity of the break positions and to values larger than 1 standard deviation ($1\sigma$). 
Conversely, when both $s_{d,0}$ and $s_{d,1}$ are allowed to vary freely, the residuals are much more flat and align well within the $1\sigma$ errors of the data. 

\begin{table*}
\begin{center}
\begin{tabular}{l ccccc}
\hline
\hline
Model  & $\delta_l/\delta/\delta_h$ & $s_{d,0}/s_{d,1}$  &  $\chi^2(\Delta \chi^2)$ & $\log(Z)$ \\
\hline
\db\ $s_{d,i}$ fixed  & $-0.42/0.58/0.35$ & $0.05/0.05$ & $582\; (-197)$ & $-359.89$ \\
\db\ $s_{d,1}$ fixed  & $-0.77/0.50/0.31$ & $0.36/0.05$ & $458\; (-73)$  & $-299.75$ \\
\db\                  & $-0.80/0.72/0.11$ & $0.30/1.43$ & $\chi^2_{\rm{ref}} = 385\; (0)$    & $-259.02$ \\
\dbin\ Model 1          & $-0.78/0.68/0.22$ & $0.26/0.82$ & $377\; (+8)$    & $-255.52$  \\
\dbin\ Model 2          & $-0.79/0.68/0.14$ & $0.31/1.26$ & $378\; (+7)$    & $-256.65$ \\
\dbin\ Model 3          & $-0.84/0.78/0.09$ & $0.28/1.57$ & $427\; (-42)$    & -281.03  \\
\hline 
& \\ 
\hline
\hline
Model  & $\delta/\delta_h$ & $s_{\rm{inj}}/s_{d,1}$  &  $\chi^2(\Delta \chi^2)$ & $\log(Z)$\\
\hline
\ib\ $s_{d,i}$ fixed  & $0.39/0.23$ & $0.05/0.05$ & $482\; (-123)$ & $-315.70$ \\
\ib\ $s_{d,1}$ fixed  & $0.40/0.24$ & $0.31/0.05$ & $371\; (-12)$  & $-263.36$ \\
\ib\                  & $0.41/0.20$ & $0.33/0.35$ & $\chi^2_{\rm{ref}} =359\; (0)$    & $-257.02$ \\

\hline
\end{tabular}
\caption{Best-fit values for slopes $\delta_l/\delta/\delta_h$ and smoothing parameters $s_{d,0}/s_{d,1}$ of the diffusion coefficient. In the case of the \ib\ model we report the value obtained for the smoothing of the injection spectrum break $s_{\rm{inj}}$ instead. The last two columns contain the $\chi^2$ and the logarithm of the Bayesian evidence, respectively. 
The $\Delta \chi^2$ is defined as $\Delta \chi^2= \chi^2_{\rm{ref}}- \chi^2$ where $\chi^2_{\rm{ref}}$ is the $\chi^2$ of reference model indicated in the table.
}
\label{tab:fitresultsD}
\end{center}
\end{table*}

As we have discussed in Sec.~\ref{sec:prod_prop}, the assumption of a homogeneous diffusion in the Milky Way is a simplification, and more realistic models, like those based on Galactic turbulence~\cite{Blasi:2012yr,Aloisio:2013tda,Aloisio:2015rsa,Evoli:2018nmb,Dundovic:2020sim} or models considering inhibited diffusion around pulsar wind nebulae~\cite{Evoli:2018aza,Fang:2019iym,Mukhopadhyay:2021dyh} predict that diffusion is inhomogeneous. To a first approximation, we expect to see effects of inhomogeneity as a $z$-dependent diffusion coefficient $D(z)$ \cite{Jacobs:2023zch}. In the following, we test if the data present a preference for models with inhomogeneous diffusion in $z$, and if also in these models there is a preference for a smooth rigidity transition of $D$.

We test the three benchmark cases dubbed \dbin\ with the $D(z)$ shape  reported in Figure \ref{fig:Dz}. For all three cases we find best-fit values for $s_{d,1}$ are much larger than 0.05 and not consistent with a broken power-law with a sharp transition of the slope.
In case of model 1 and model 2 we obtain a slightly better $\chi^2$ with respect to the homogeneous case ($\Delta \chi^2=8$) and with a value of $s_{d,1}$ which is $0.82^{+0.11}_{-0.22}$ for Model 1 and $1.26^{+0.20}_{-0.42}$ for Model 2, i.e.~they are consistent at $1\sigma$ errors. The smoothing of the low-energy break is very similar to the one obtained for the homogeneous $D$ model. This implies that a very smooth function of rigidity is required for $D$  also when more physical models for $D(z)$ are assumed. The best-fit values obtained for $\delta_l/\delta/\delta_h$ with model 1 and model 2 are very similar to the ones obtained with \db.
For Model 3, which has a significantly different trend of $D(z)$, we obtain a worse goodness of fit ($\Delta \chi^2=-42$). Although this case provides still an overall good fit to the data, it seems disfavored with respect to the other three. This is confirmed by the value of the natural logarithm of the ratio of the Bayesian evidence $Z$ of Model 3 and \db\ which is about 22.
Nonetheless, also with this very different shape for the diffusion coefficient in $z$ we find a very significant smoothing with both low energy and high energy $s_{d,i}$ much larger than 0.05.
Finally, the overall $\chi^2$ and Bayesian evidence for the \db\ model  and \dbin\ Model 1 and 2 are very similar indicating that there is no clear preference for either of the models given the data currently available.

In Fig.~\ref{fig:Dz}, we present the best-fit profiles of $D(z)$ as a function of $z$ at a fixed rigidity of $R=4$ GV. The curves predict different diffusion for $|z|<0.5$ kpc, near the Galactic disk. However, as we move to regions further away from the Galactic plane, all four cases converge to similar values of approximately $2\times 10^{28} -4 \times 10^{28}$ cm$^2$/s. This result can be explained in terms of the residence time. CRs spend a considerably longer time in the diffusive halo compared to the Galactic disk. Consequently, the diffusion profiles tend to converge in regions where CRs predominantly reside within the halo. This fact indicates that current CR data, considering also the similar $\chi^2$ values obtained for the different models with convection, are not able constrain the shape of the diffusion coefficient with $z$.

Considering the \ib\ model the results are not as clear for the \db\ model. While there is still a preference for a smooth high-energy break in diffusion, it is now observed with less smoothing and a smaller significance. The best-fit value of $s_{D,1}$ is $0.35\pm0.08$, and the difference in $\chi^2$ compared to the case with a sharp transition at the break is 12, corresponding to a significance of $3.5\sigma$. On the other hand, the smoothing in the low-energy  break in the injection spectrum present in this model, is highly significant, with a $\Delta \chi^2$ of 122.
For this model, we have not tested any inhomogeneity of the diffusion coefficient because, following the results obtained with the \db\ models, we do not expect any significant change in the shape of $\delta(R)$ if we include a specific function for $D(z)$.
Moreover, we remind that the \ib\ model is disfavored by the too-extreme values required for some  nuclear cross sections and the too-high secondary $e^+$ flux below 2 GeV.

A possible way to investigate the shape of $D$ in the inner part of the Galaxy, close to the disk, is using $^{10}$Be flux data. In fact, this isotope has a decay time of about 2 Myr. In particular, the data of $^{10}$Be/$^9$Be and Be/B ratio has been demonstrated to be able to rule our very small values of $L$ (see, e.g., \cite{Evoli:2019iih,Cuoco:2019kuu,Maurin:2022gfm,DiMauro:2023oqx,Jacobs:2023zch}). In the future, the data of AMS-02 and the balloon payload experiment HELIX on $^{10}$Be/$^9$Be and Be/B could provide important information about the shape in $z$ if the diffusion coefficient. 
In particular, HELIX is designed to measure, with unprecedented precision the $^{10}$Be/$^9$Be ratio below 1 GeV/nuc \cite{Wakely:2023/r}.

\begin{figure}
  \centering
\includegraphics[width=0.49\textwidth]{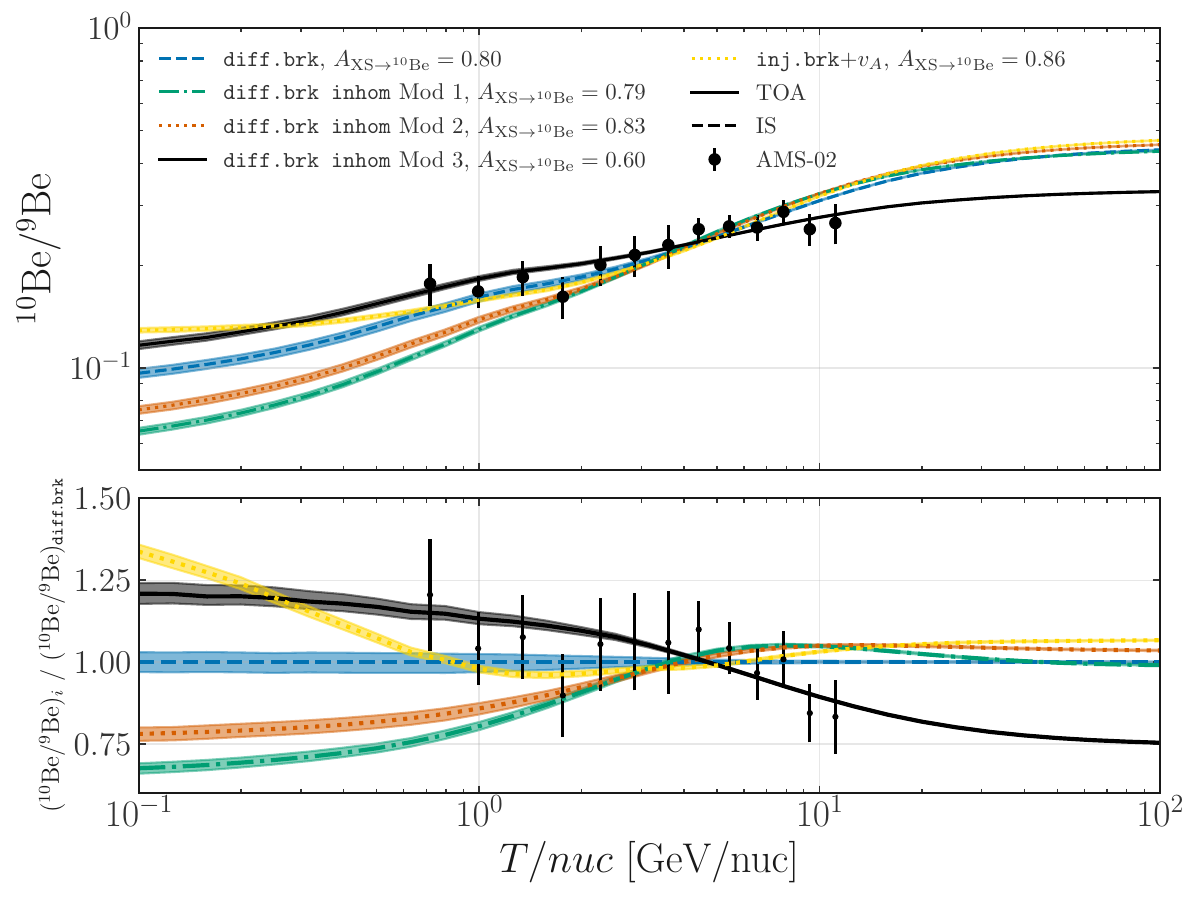}%
  \caption{
Comparison of the preliminary AMS-02 data for $^{10}Be$/$^9$Be \cite{AMSBeICRC2023} with the theoretical predictions of the various models considered in this work evaluated at their  best-fit. We also report, for each model, the $^{10}$Be production cross-section renormalization factor which has been applied in order to achieve a better match to the data. }
\label{fig:isotopes}
\end{figure}

In order to demonstrate the potential of the Be isotope data, we consider preliminary AMS-02 data for $^{10}$Be/$^9$Be \cite{AMSBeICRC2023}, and we compare them with the theoretical predictions of the various models considered here evaluated at the best-fit of each model.
We show this result in Figure~\ref{fig:isotopes}.

In order to have a better match with the data we have performed a slight renormalization of the $^{10}$Be flux with factors of order 1, specifically by a factor of 0.80/0.79/0.83/0.60/0.86 for the \db\ / \dbin\/ model 1/ 2/ 3 and \ib model. These rescaling factors can be reabsorbed within the uncertainties of the production cross section of $^{10}$Be, which is still poorly known \cite{Maurin:2022gfm}. We also remind that this isotope contributes by a factor of about 10\% to the total yield of Be. Therefore, a small rescaling of the $^{10}$Be cross sections can be compensated with a very minor rescaling in the opposite direction for the other isotopes, without changing the fit to the other CR data.
We note that all the models require similar renormalization factors for the $^{10}$Be cross section of the order of 0.8.
This is slightly different for the other species which require either no changes or small increases of the order of 5-10\%. Nonetheless, these rescaling are well within the  large uncertainties in $^{10}$Be cross sections \cite{PhysRevC.98.034611}.  
This also implies that future  precise measurements, with errors below $20\%$, of the cross section for the $^{10}$Be isotope could help in probing the inhomogeneity of the diffusion coefficient.

The resulting $\chi^2$ for the $^{10}$Be/$^{9}$Be data amount to 10/27/22/7/13 for the \db\ / \dbin\/ Mod 1/ 2/ 3 and \ib 
 models respectively. Therefore, Model 3 is the one that best fits the Be isotopes. At the same time, however, this model is also the one that provides the worst $\chi^2$ in the fit to all the other CRs (see Table \ref{tab:fitresultsD}). 
In particular, it seems to fit poorly the Be/C data, as can be seen from the Be/C residuals shown in the Appendix for the various models.
Indeed, the $\chi^2$ for the fit to the Be/C ratio obtained with \dbin\ Model 3 is about $\sim 60$ while the other models have values around 30.
The propagation setups \db\ and \ib\ provide a decent fit to the $^{10}$Be/$^{9}$Be data with reasonable values for the cross sections. These two models are able to reproduce well and at the same time the lower and higher energy part of the AMS-02 data. The other two cases, \dbin\ Model 1 and 2, can explain well the ratio at intermediate energies, but over-estimate the data at high energies, and under-estimate them at low energies.
The situation about all the \dbin\ models thus seems inconclusive.
Nonetheless, this demonstrates that precise $^{10}$Be/$^{9}$Be data can be a sensitive probe of inhomogeneous diffusion scenarios.

Future AMS-02 and HELIX data at kinetic energies per nucleon in the range between 0.1 to a few tens of GeV/nuc seem to be able to discriminate between different models of $D(z)$.
This is particularly true  at 0.1 GeV/nuc where the models
have different predictions  at the level of $50\%$ (see Figure \ref{fig:isotopes}). 

Further progress in understanding inhomogeneous diffusion may be expected through the combination of CR data with other messengers such as gamma-rays or radio observations. These additional measurements provide information from other locations in the Galaxy and thus have the potential to provide more conclusive evidence and enable a better characterization of the diffusion processes in the Galaxy.\\

\section{Conclusions}
\label{sec:conclusions}

In this paper, we have used the latest CR data from Voyager, AMS-02, CALET and DAMPE for species from proton up to Oxygen to investigate the shape of the diffusion coefficient in rigidity and as a function of $z$ the distance from the Galactic plane.
We try two different propagation setups. The first, called \db\ (and \dbin\ in the inhomogeneous case),  has convection, two breaks in the rigidity dependence of the diffusion coefficient, and a single power law  for the injection spectrum of primary CRs, with different slopes for p, He and CNO.
The second \ib\ parametrizes both the injection spectra of primary CRs and  diffusion with a single break broken power law. This model does not include convection which is replaced with  reacceleration. We find very good fits to the CR data for both models with $\tilde{\chi}^2$ of the order of 0.5.

We have demonstrated that for the \db\ and \dbin\ propagation setups the rigidity dependence of the diffusion coefficient exhibits two breaks at $\sim$ 4 GV and $\sim$ 250 GV. For the first time, we provided very strong evidence (with more than $8\sigma$ significance) that these two breaks are smooth in shape with smoothing coefficients of the order of $s_{d,0}=0.3$ and $s_{d,1}=0.8-1.4$.
The smoothness of the breaks is required in particular to achieve a good fit to the spectra of primary CRs, especially protons and helium.
We find this result to be true both for homogeneous and inhomogeneous diffusion models.
In particular, we show that the shape of the diffusion slope $\delta=dD/dR$ is very close to the one derived from models based on self-generated turbulence \cite{Blasi:2012yr,Aloisio:2013tda,Aloisio:2015rsa,Evoli:2018nmb,Dundovic:2020sim}.

When using the \ib\ propagation setup, instead, the evidence for a smooth transition of the diffusion coefficient at high energy is less evident ($\sim 3.5\sigma$ significance). However, the values of the nuclear cross sections required to fit the CR data in this setup stretch the current known uncertainties for these quantities, in particular in the secondary Carbon case.
Furthermore, we have shown that the prediction for the secondary $e^+$ in the \ib\ setup is well above the AMS-02 data for energies below 2 GeV.

We have compared the models studied in this paper with preliminary data of $^{10}$Be/$^{9}$Be ratio. These measurements indicate a preference for the \dbin\ Model 3.  On the other hand, this same model is disfavored in the fit with all the other CRs.
Therefore, at the moment none of the tested models is significantly preferred over the others.
This conclusion might change with future precise data from AMS-02 and HELIX for the ratio between $^{10}$Be/$^{9}$Be CRs.
which will probably be able to shed light on a possible inhomogeneous dependence of the diffusion coefficient in the Galaxy. The kinetic energies at around 0.1 GeV/nuc are particularly interesting because, at these energies, the predicted fluxes of the models we tested are very different from each other.

\begin{acknowledgments}
The authors thank Yoann Genolini, Silvia Manconi and Luca Orusa for the comments provided on the draft.
M.D.M. and A.C.~acknowledge support from the Research grant {\sl TAsP (Theoretical Astroparticle Physics)} funded by Istituto Nazionale di Fisica Nucleare (INFN). \\
A.C. acknowledges support from: {\sl Departments of Excellence} grant awarded by the Italian Ministry of Education, University and Research (MIUR); Research grant {\sl The Dark Universe: A Synergic Multimessenger Approach}, Grant No. 2017X7X85K funded by the Italian Ministry of Education, University and Research (MIUR); 
 Research grant {\sl Addressing systematic uncertainties in searches for dark matter}, Grant No. 2022F2843L funded by the Italian Ministry of Education, University and Research (MIUR);
 {\sl Research grant The Anisotropic Dark Universe}, Grant No. CSTO161409, funded by Compagnia di Sanpaolo and University of Torino. \\
M.K. is supported by the Swedish Research Council under contracts 2019-05135 and 2022-04283 and the European Research Council under grant 742104.

\end{acknowledgments}

\bibliographystyle{apsrev4-1}
\bibliography{paper}

\clearpage
\newpage

\appendix

\section{Results for the propagation and cross-section parameters}
\label{sec:app}

\begin{figure*}
  \centering
\includegraphics[width=0.5\textwidth]{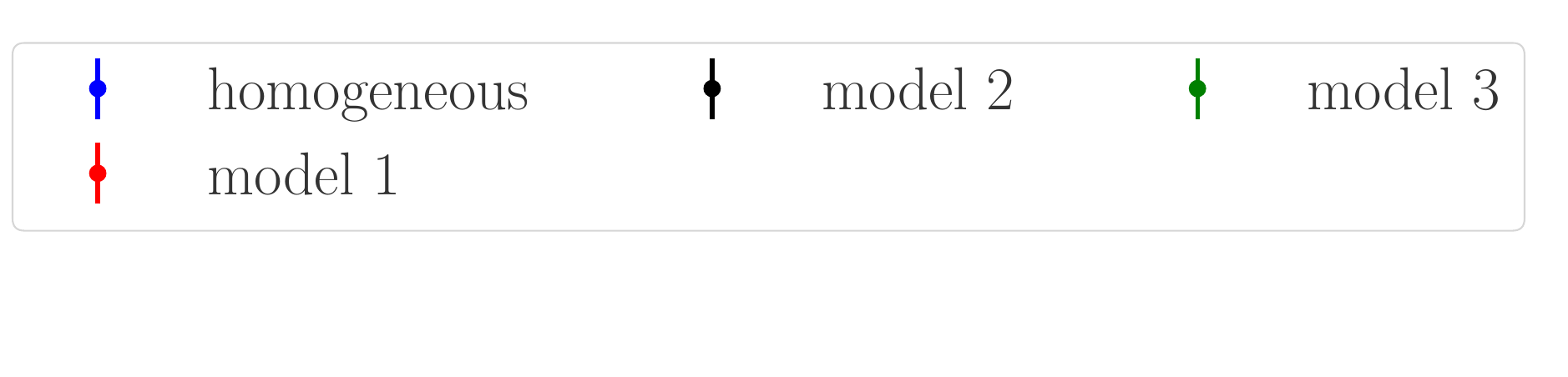}
\includegraphics[width=0.5\textwidth]{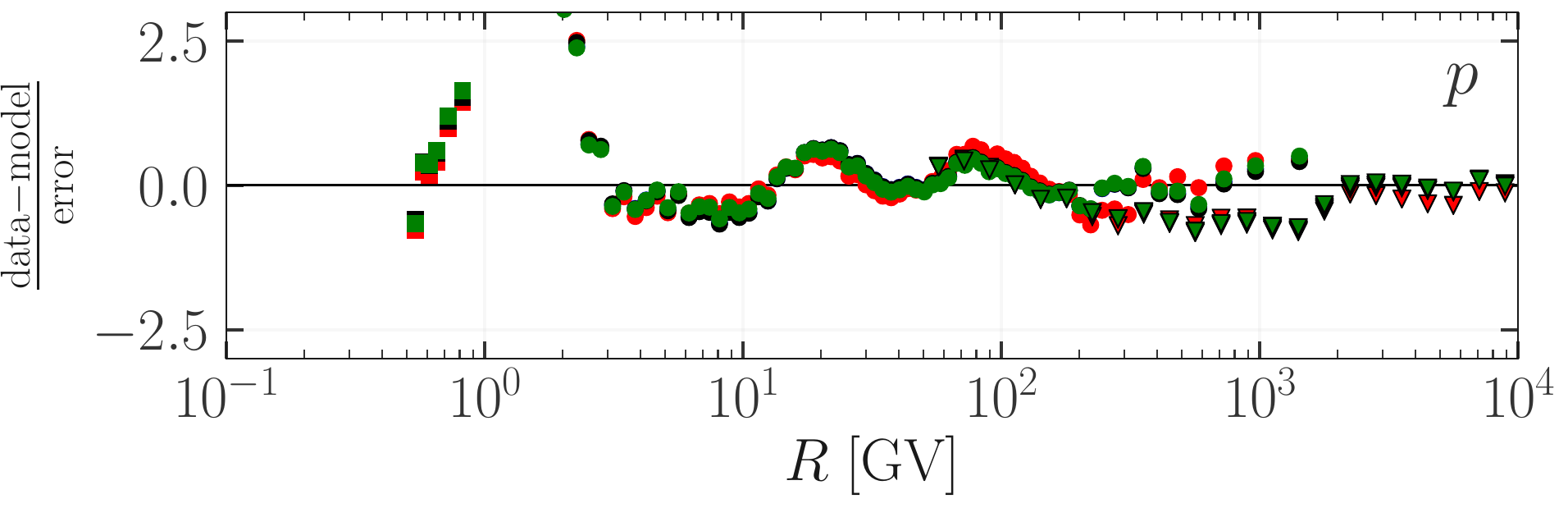}%
\includegraphics[width=0.5\textwidth]{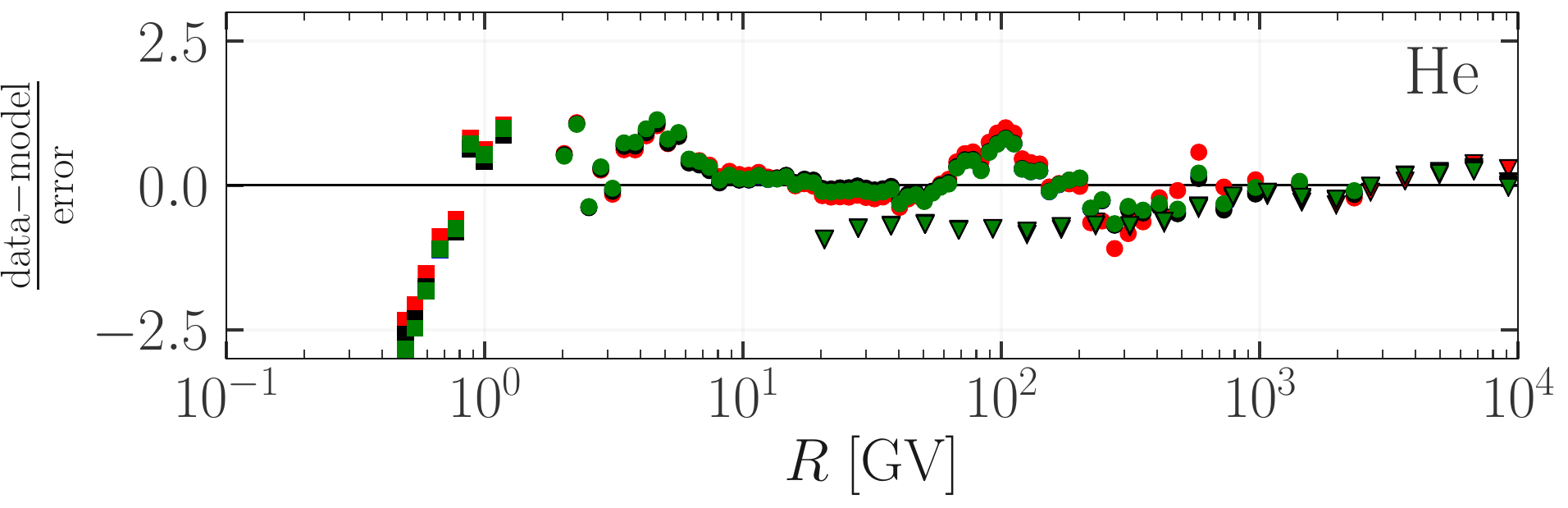}
\includegraphics[width=0.5\textwidth]{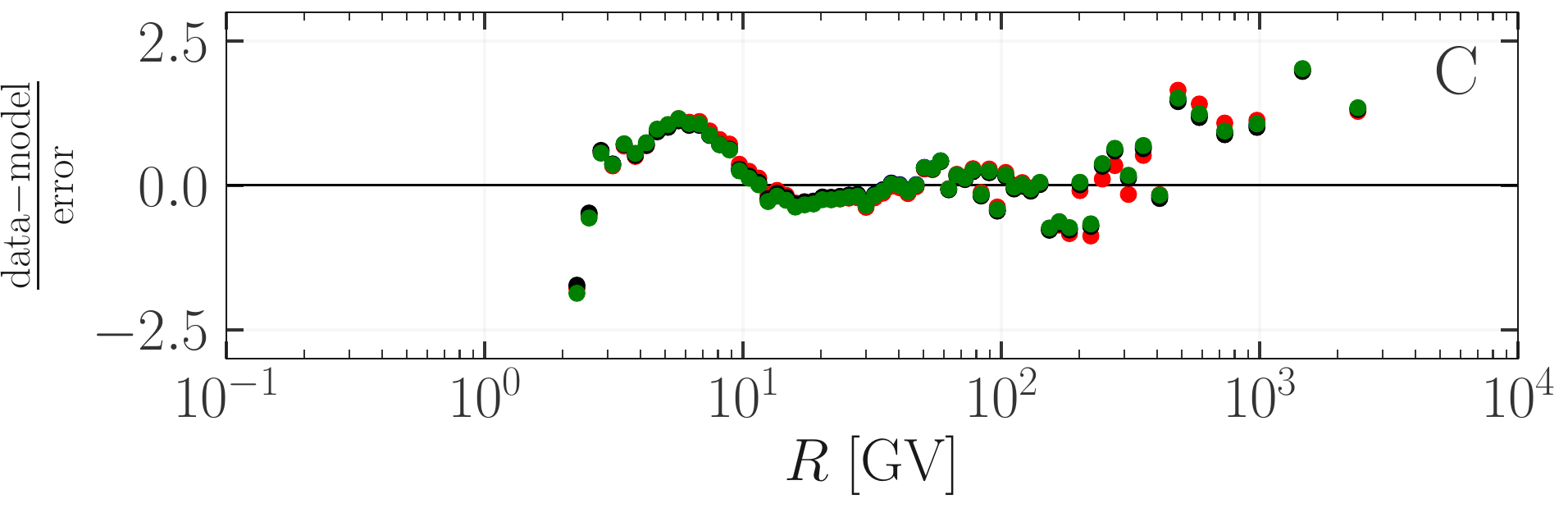}%
\includegraphics[width=0.5\textwidth]{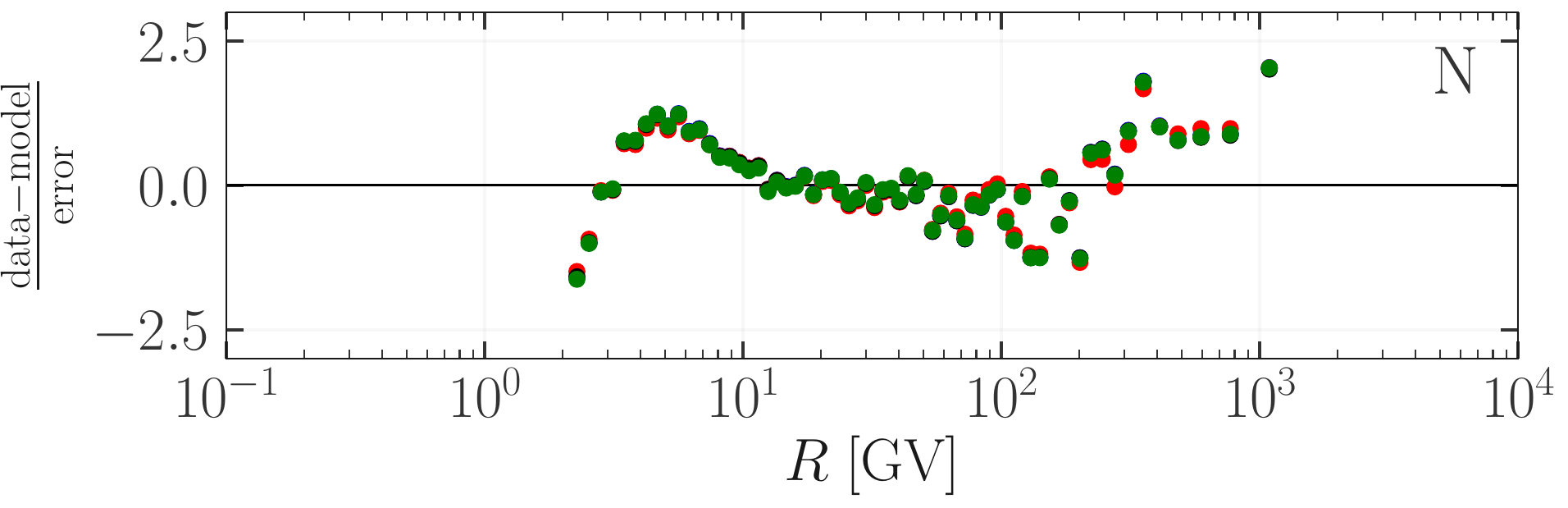}
\includegraphics[width=0.5\textwidth]{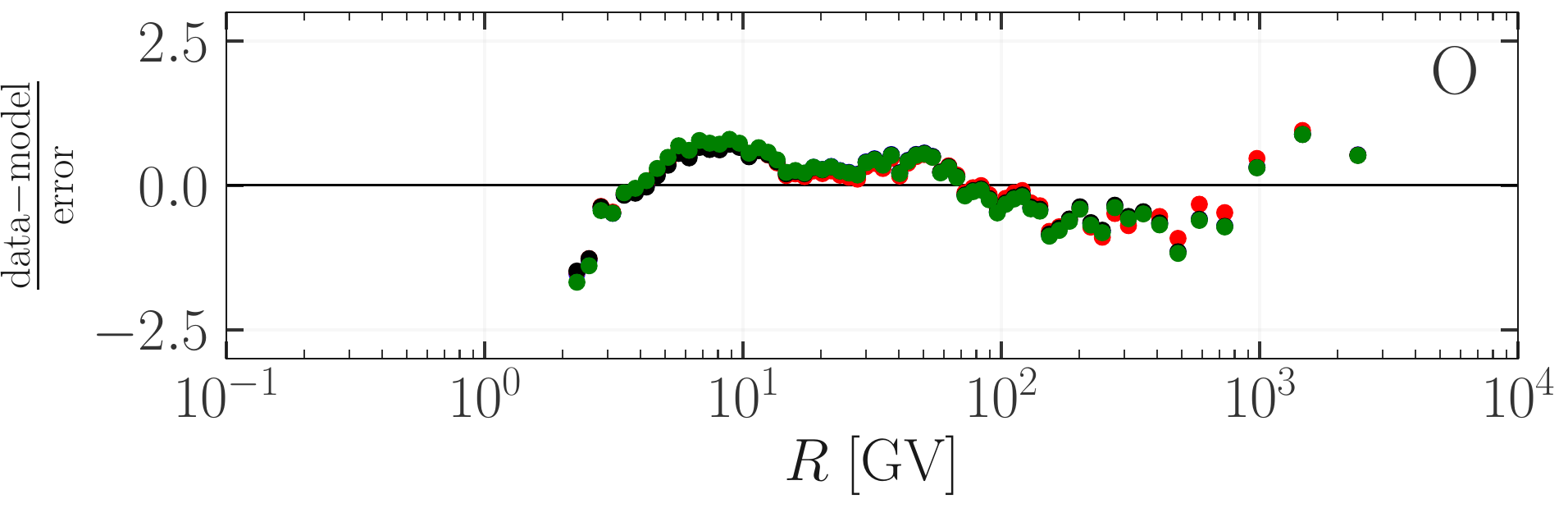}%
\includegraphics[width=0.5\textwidth]{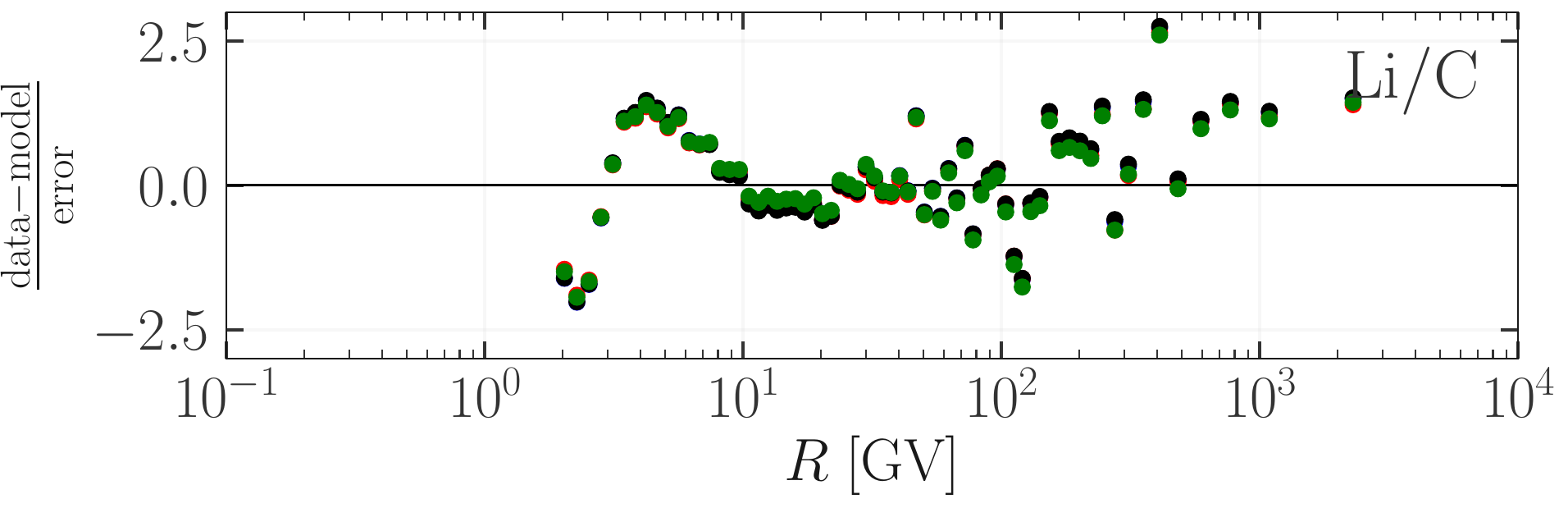}
\includegraphics[width=0.5\textwidth]{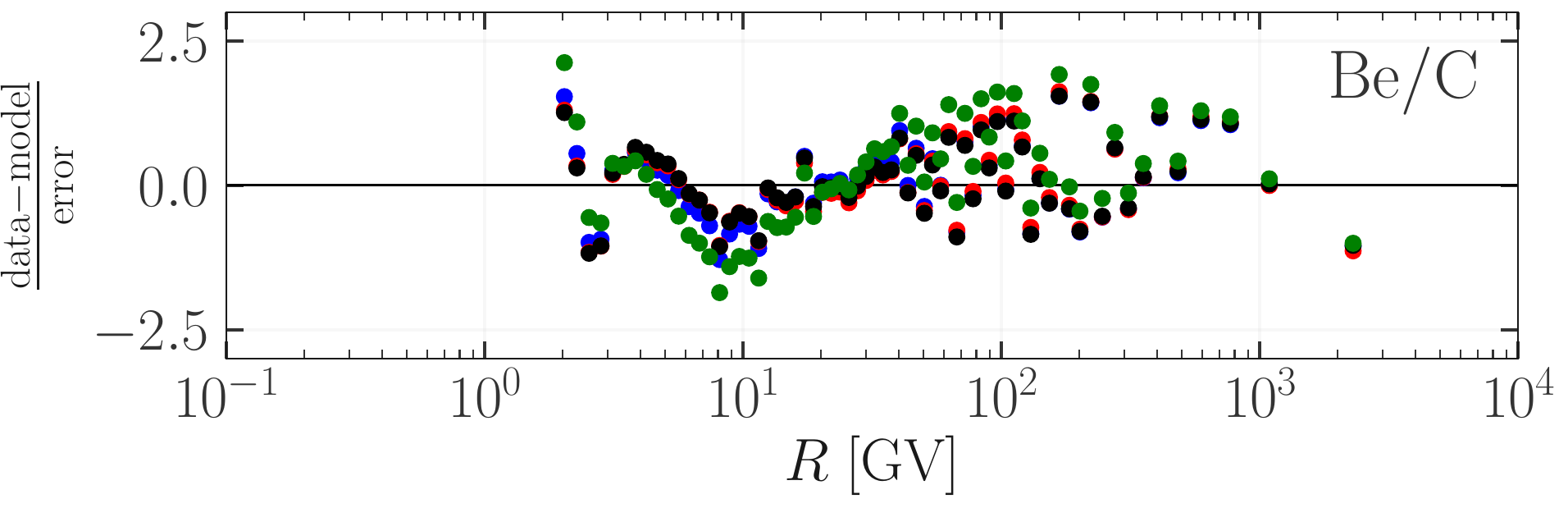}%
\includegraphics[width=0.5\textwidth]{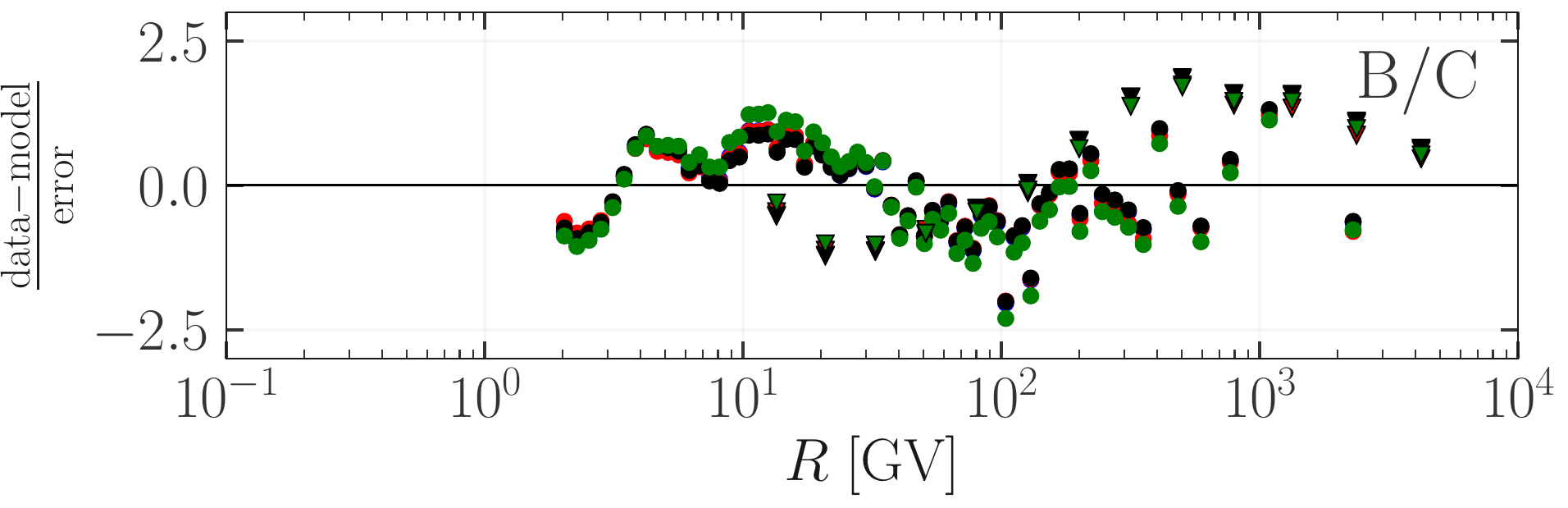}
\includegraphics[width=0.5\textwidth]{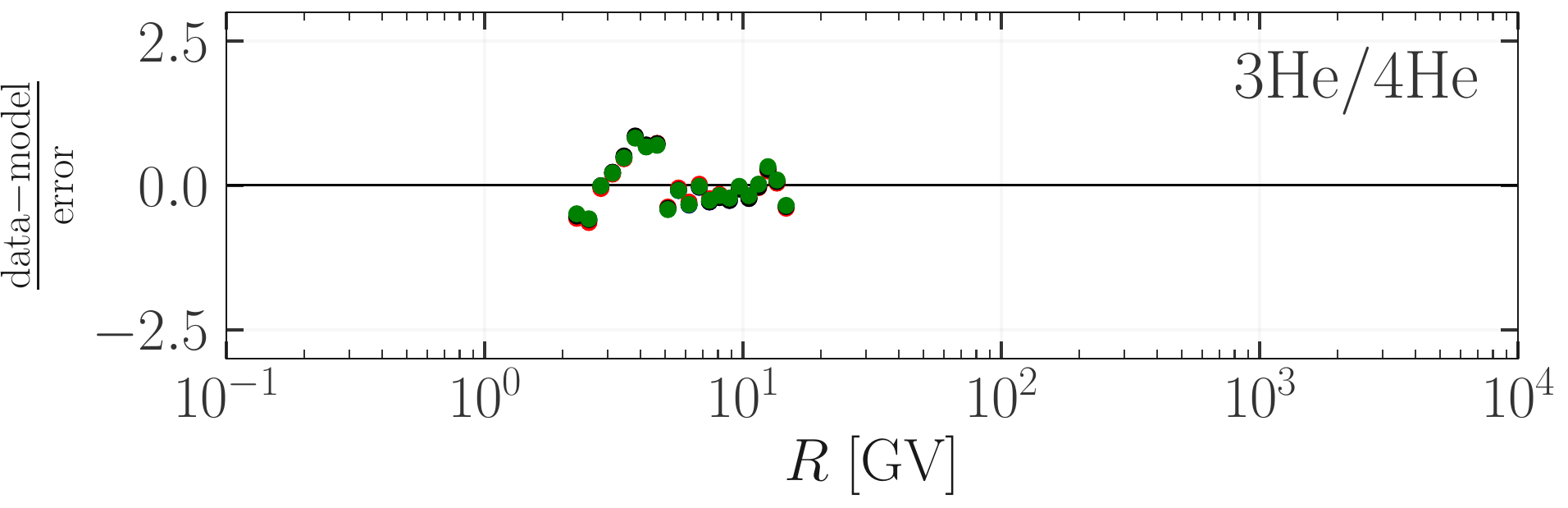}%
\includegraphics[width=0.5\textwidth]{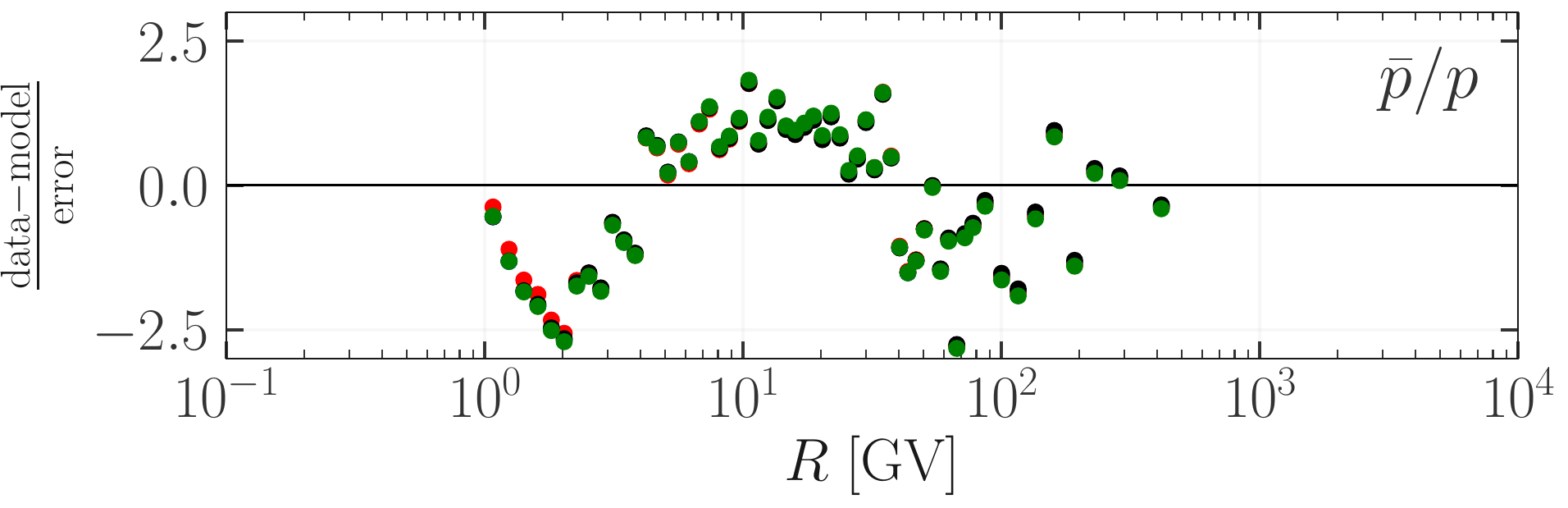}
  \caption{
     Residuals reported as $(\rm{data}-\rm{model})/\rm{error}$ for the \db\ and \dbin\ models considered in the paper. We show the residuals obtained for all the CR data included in the analysis.
  }
  \label{fig:CRresidualsALLdb}
\end{figure*}

\begin{figure*}
  \centering
\includegraphics[width=0.5\textwidth]{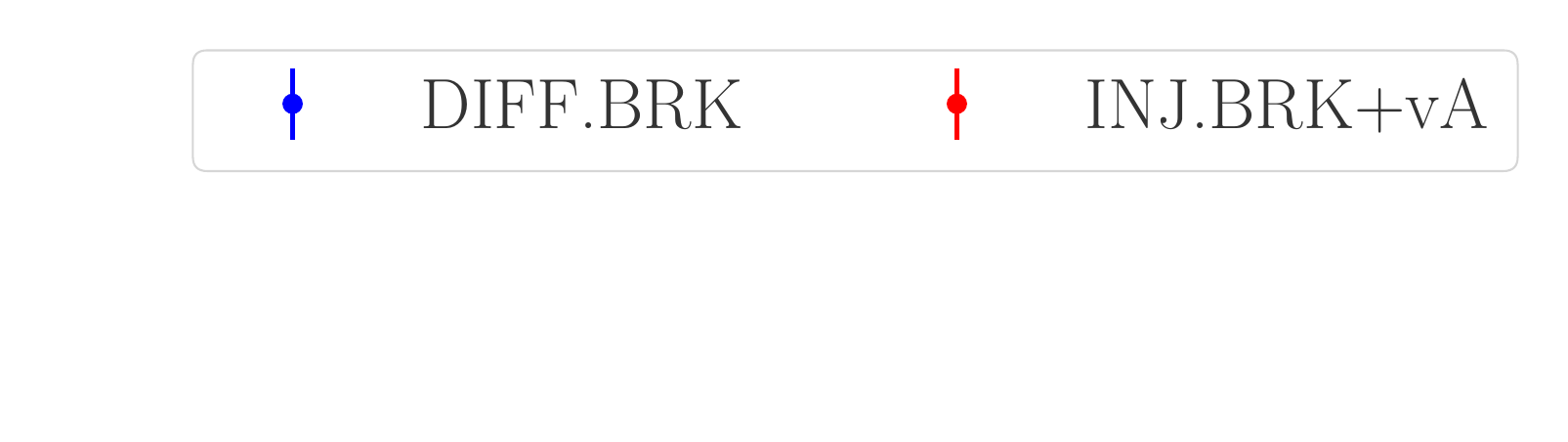}
\includegraphics[width=0.5\textwidth]{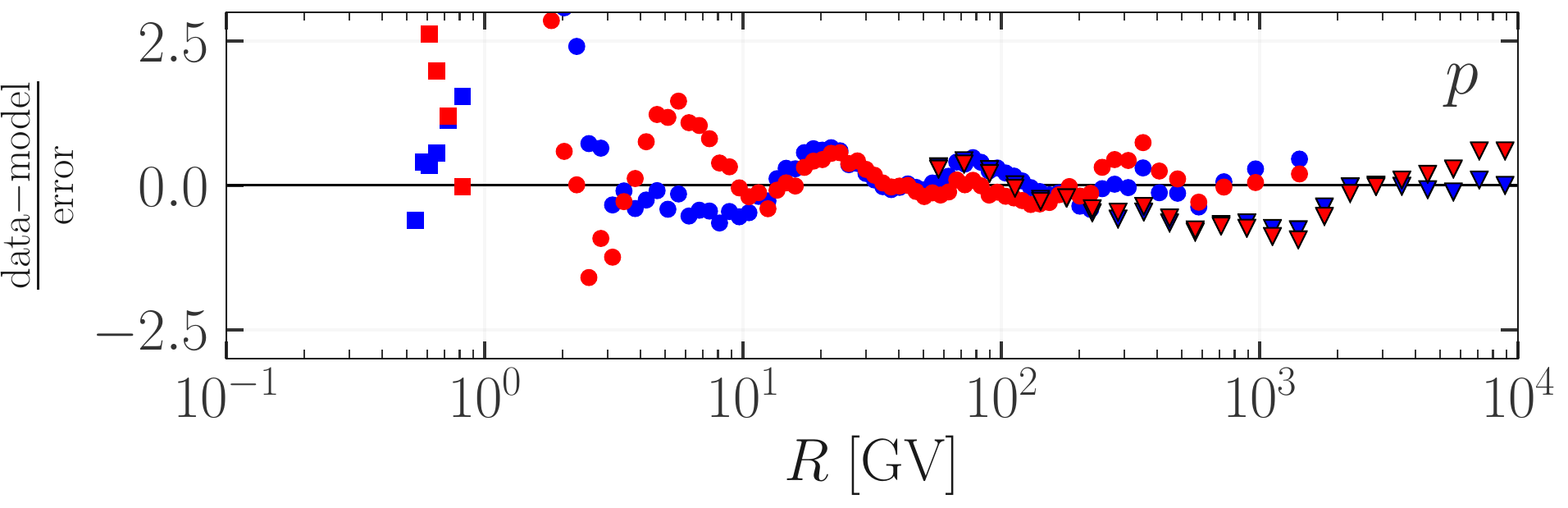}%
\includegraphics[width=0.5\textwidth]{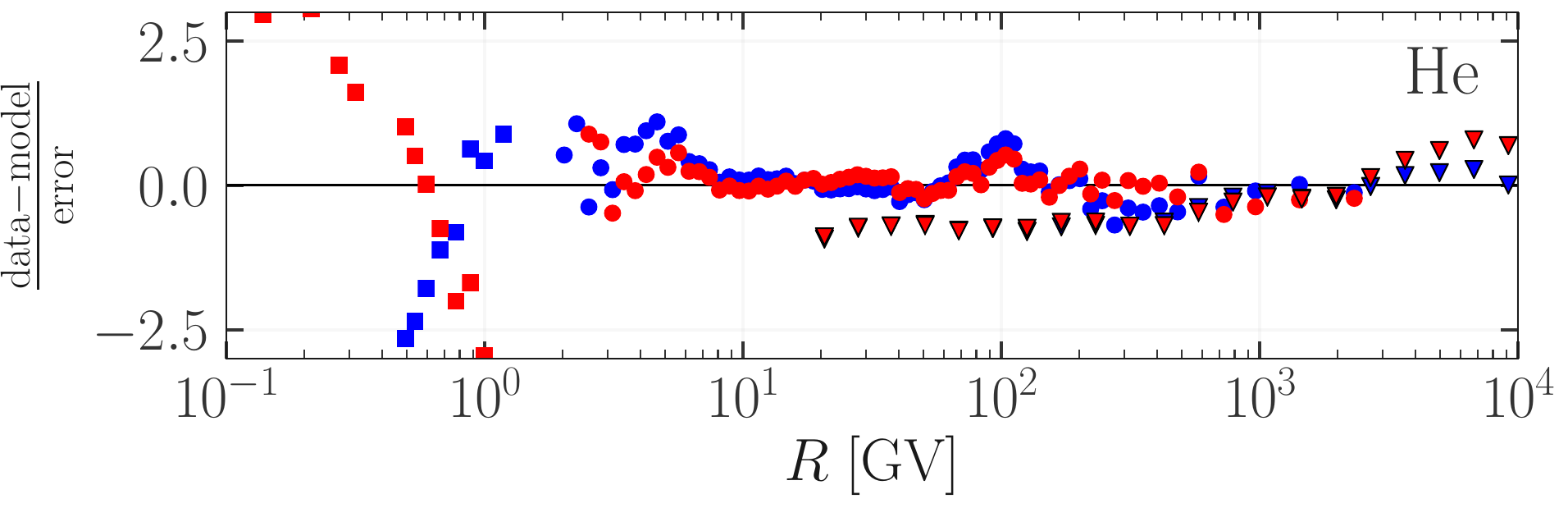}
\includegraphics[width=0.5\textwidth]{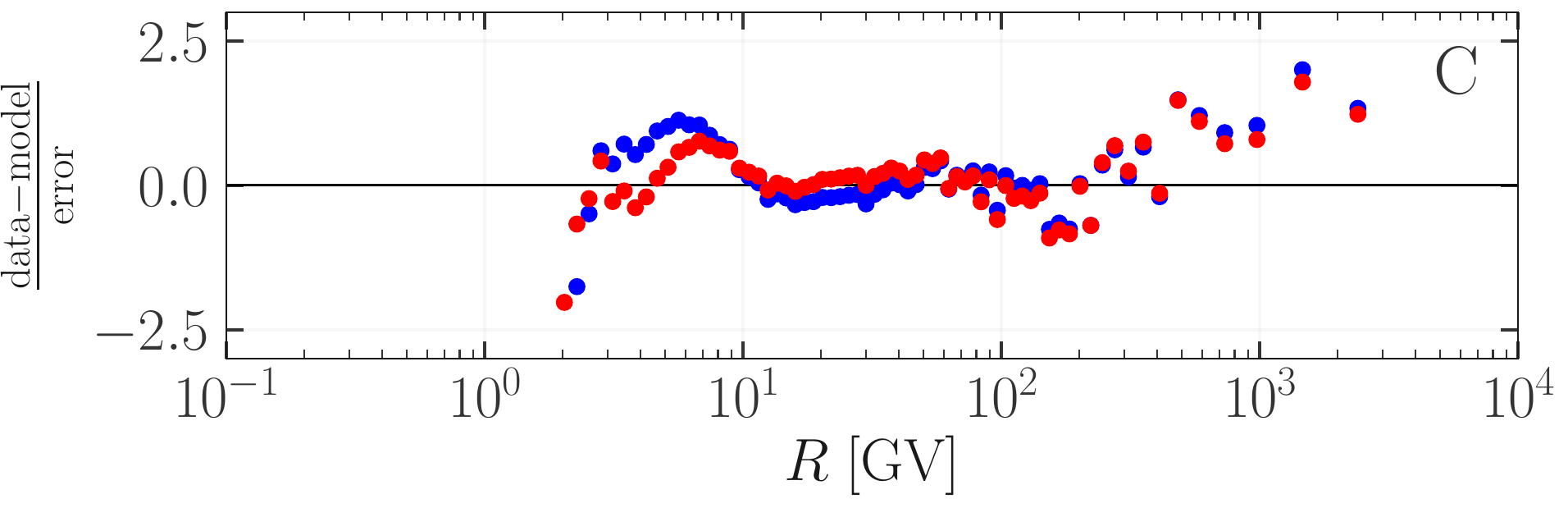}%
\includegraphics[width=0.5\textwidth]{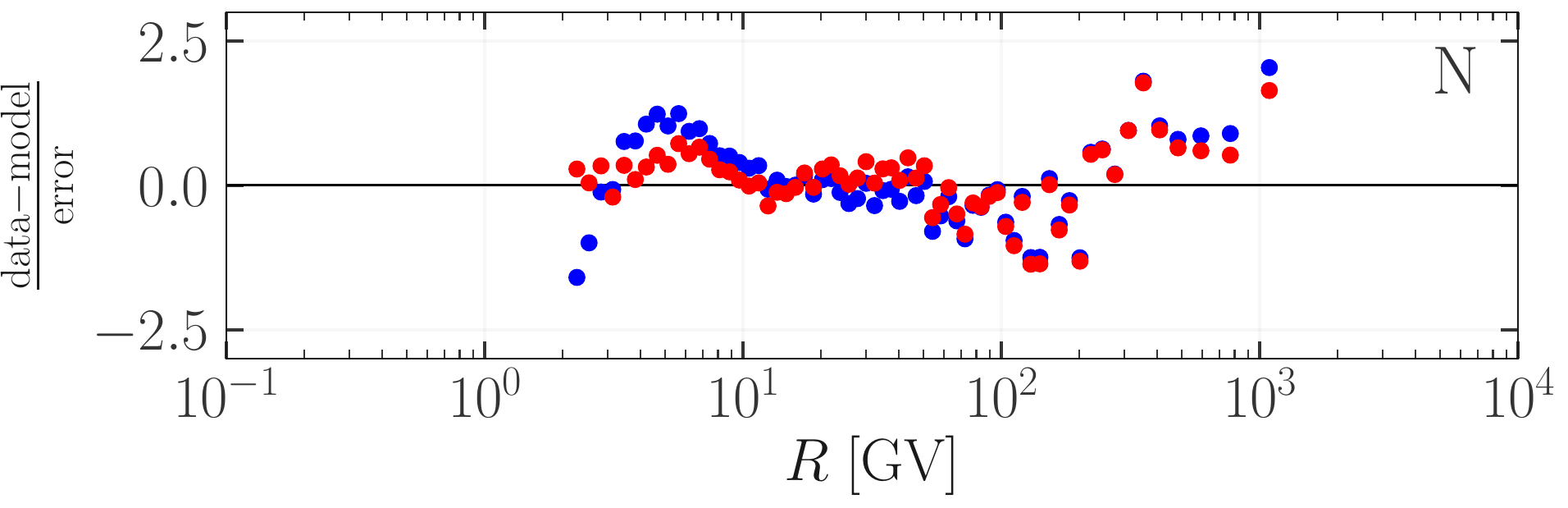}
\includegraphics[width=0.5\textwidth]{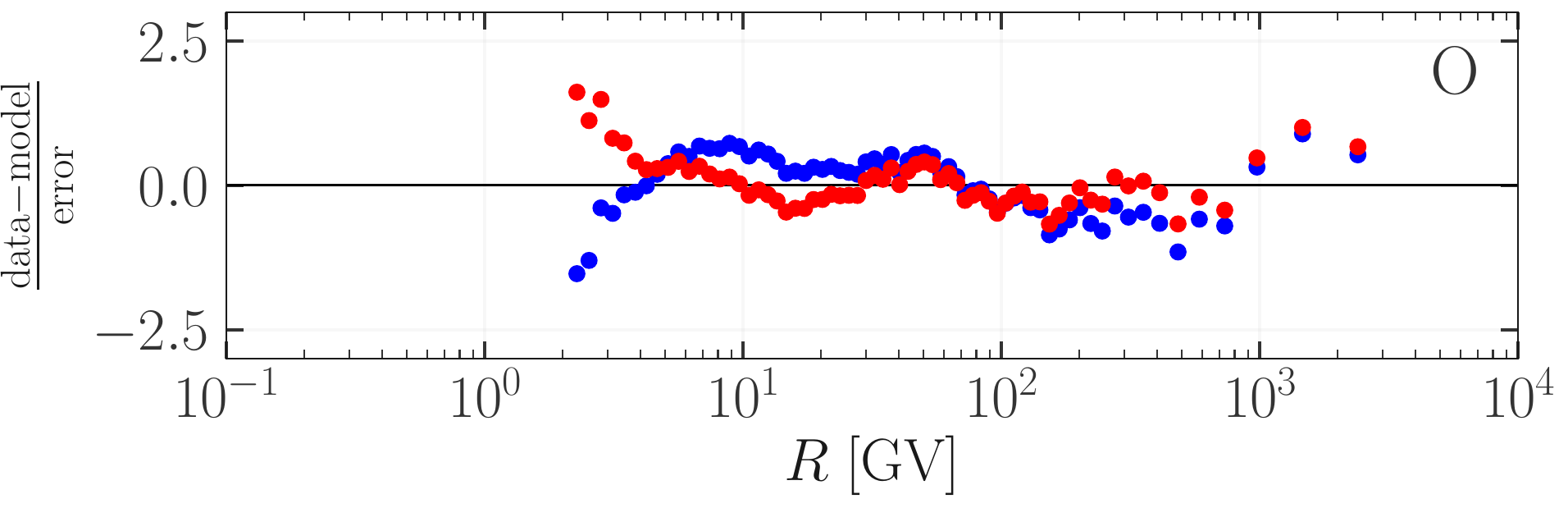}%
\includegraphics[width=0.5\textwidth]{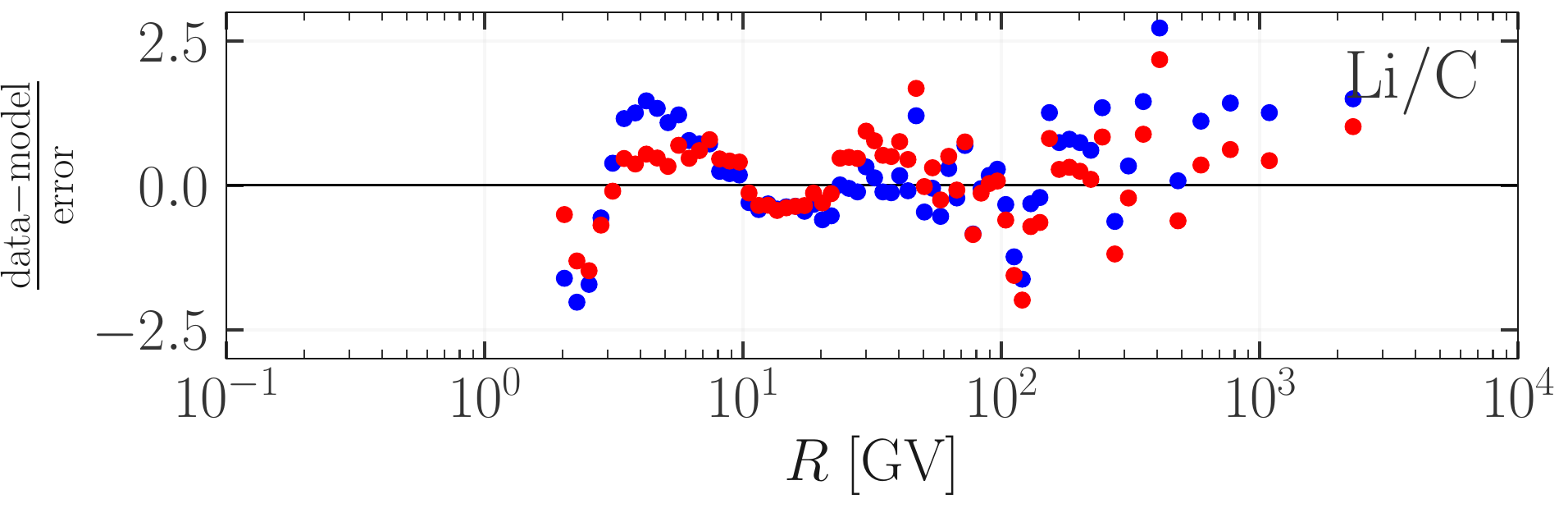}
\includegraphics[width=0.5\textwidth]{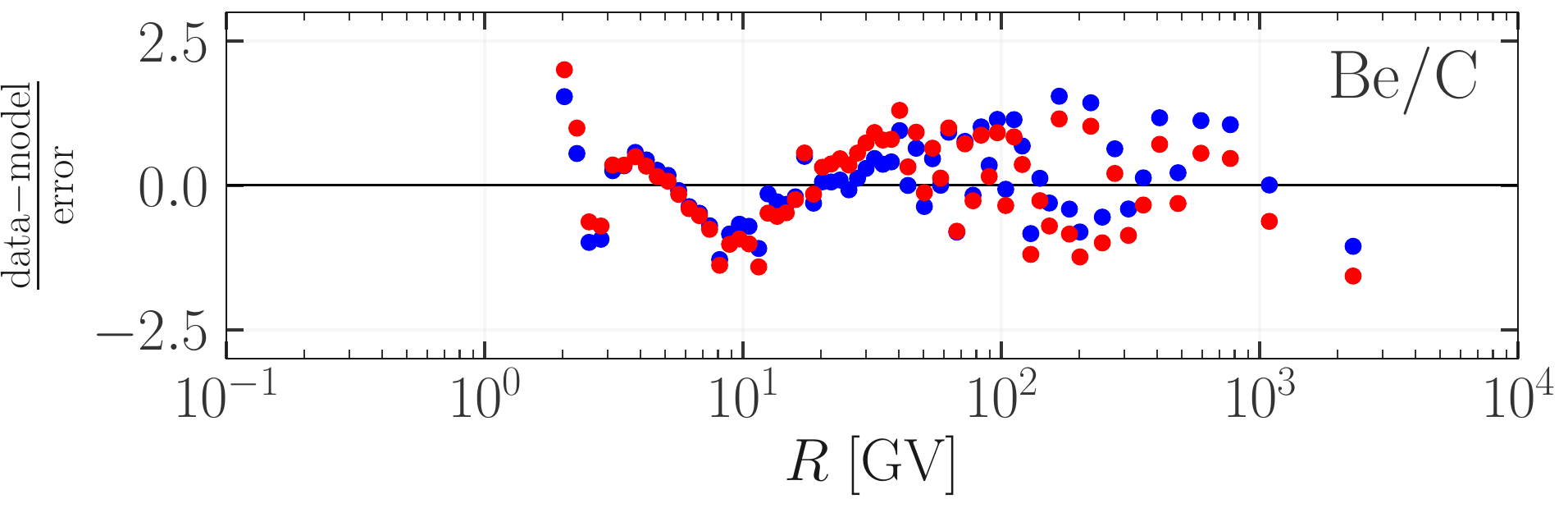}%
\includegraphics[width=0.5\textwidth]{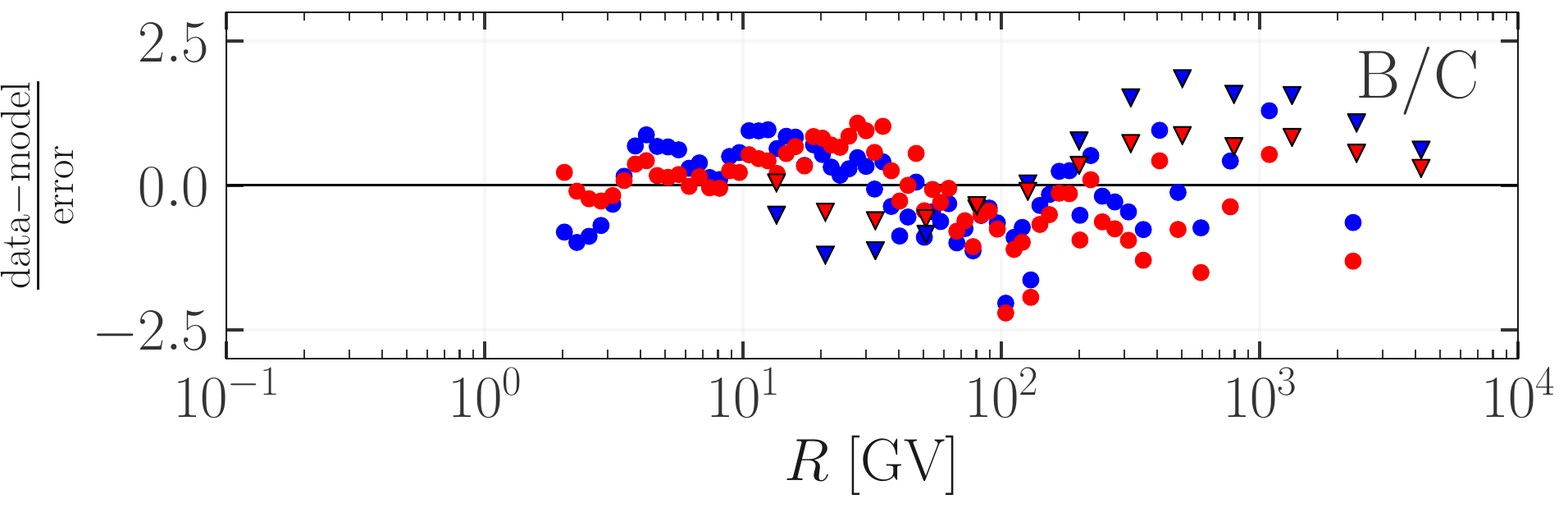}
\includegraphics[width=0.5\textwidth]{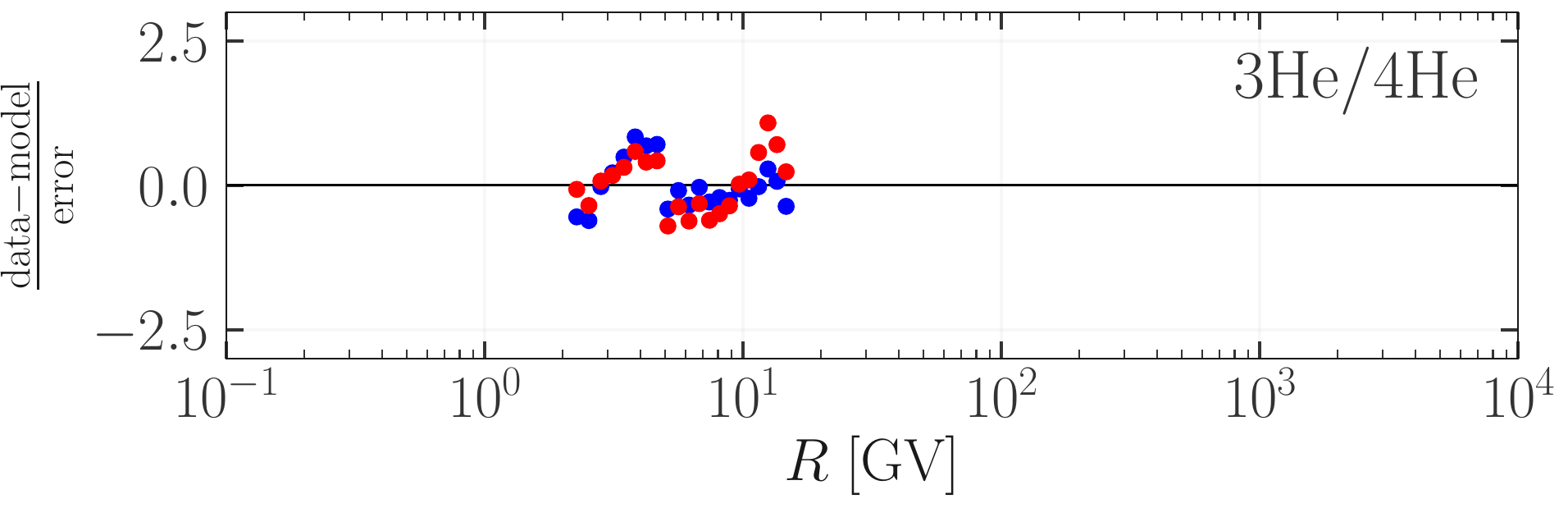}%
\includegraphics[width=0.5\textwidth]{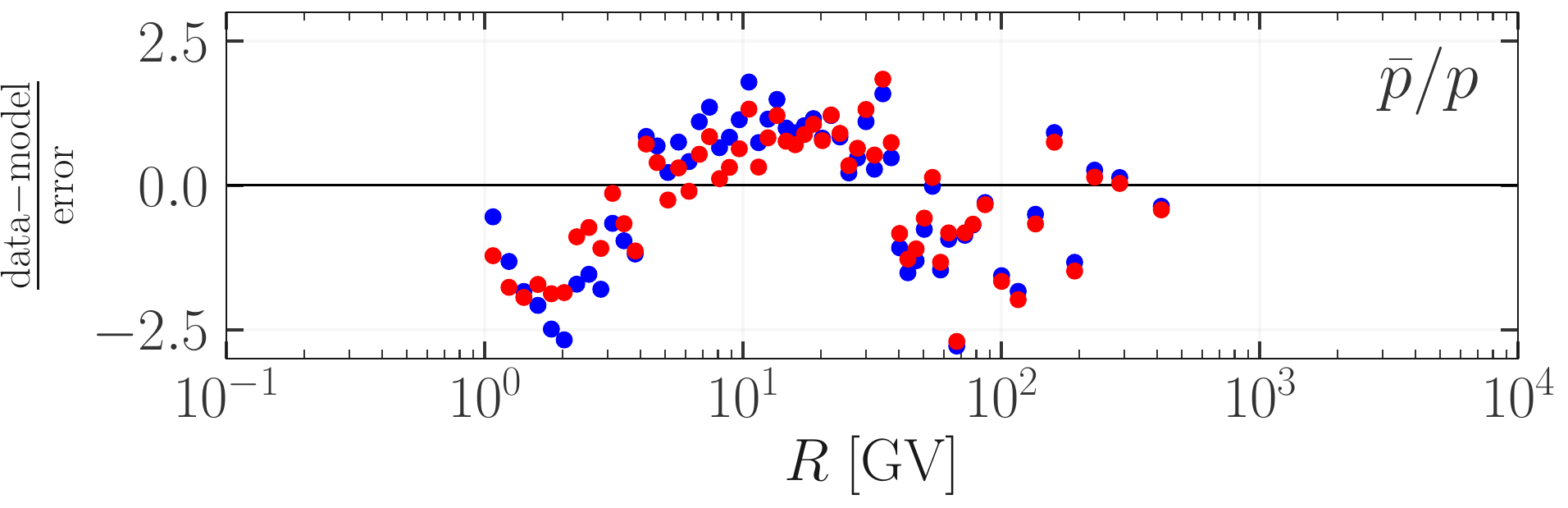}
  \caption{
    Same as Fig.~\ref{fig:CRresidualsALLdb} for the \db\ and \ib\ cases.
  }
  \label{fig:CRresidualsALLinjb}
\end{figure*}

In this section, we report further results for the propagation and cross-section parameters as well as some figures with the fit to the CR data. In Fig.~\ref{fig:CRresidualsALLdb} and \ref{fig:CRresidualsALLinjb} we show the residuals of the model fit relative to the CR data, for the various models considered in the paper. In particular, we can see that \dbin\ Model 3 is the one that provides the worst fit to the Be/C secondary over primary flux ratios. The fit to primary CR flux data is instead very similar for all the models with convection tested. On the other hand, we see that the model \ib\ fits better the primary CR data while the secondary over primary measurements are explained with a similar goodness of fit for the \db\ and \ib\ cases.

In Tab.~\ref{tab:full_results_db} and \ref{tab:full_results_injb} we show the best fit and errors for all the parameters of the models.

\begin{figure*}[t]
\centering
\includegraphics[width=0.33\textwidth]{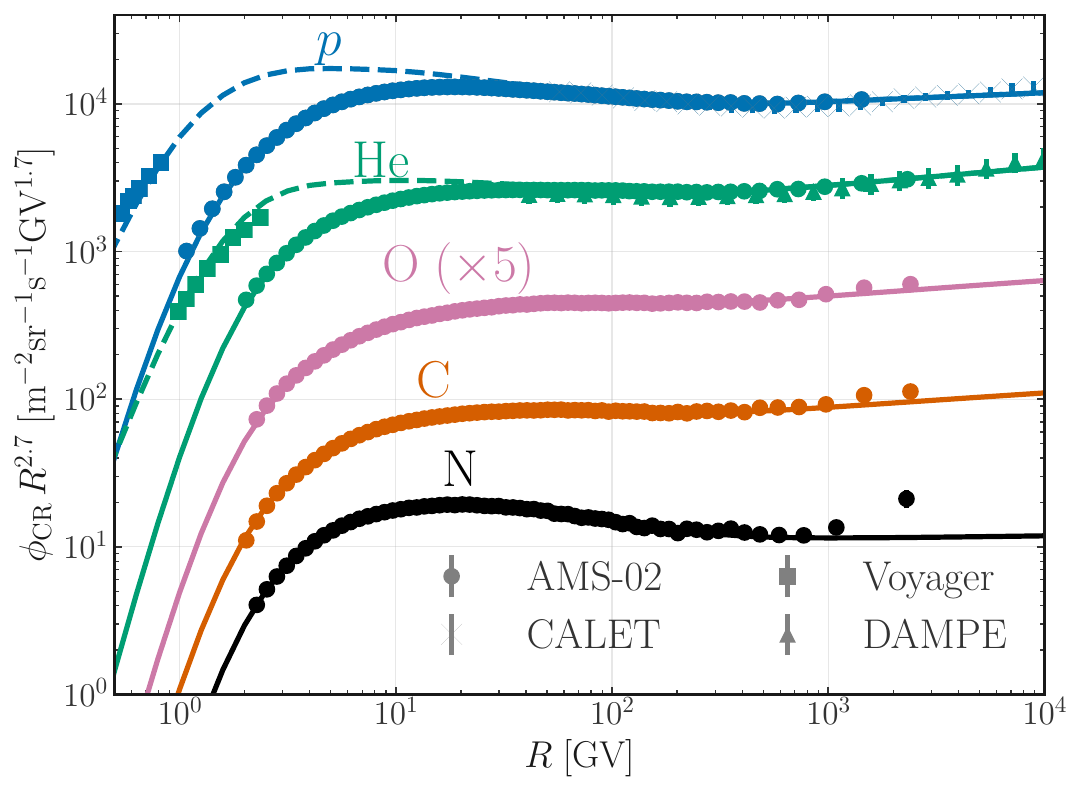}%
\includegraphics[width=0.33\textwidth]{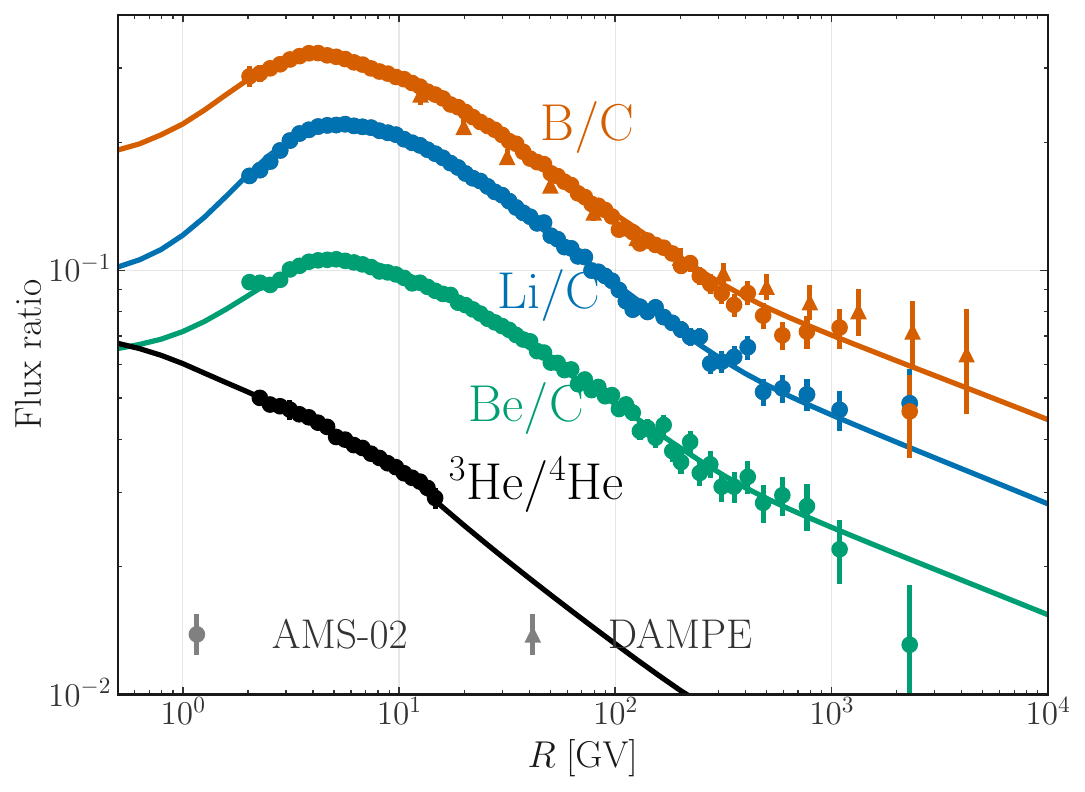}%
\includegraphics[width=0.33\textwidth]{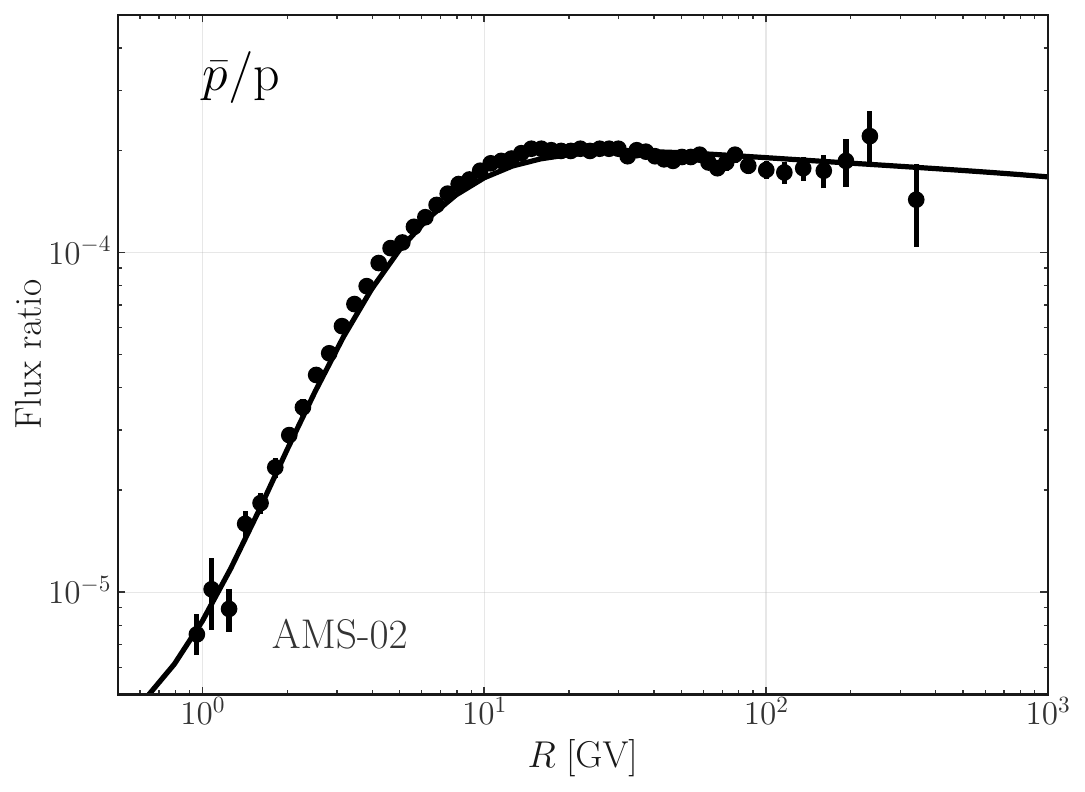}
\caption{
   Same as Fig.~\ref{fig:CRfit}, but for the \ib\ model. 
  }
  \label{fig:CRfit_reacc}
\end{figure*}

\begin{table*}
    \caption{Summary of the best-fit and error values for the parameters of the models that we fit to the data. In the rows that contain the $\chi^2$ of the fit to each individual dataset we report in brackets the number of datapoints.
    }
    \label{tab:full_results_db}
    \begin{tabular}{l rcl c c c c c c c }
\hline \hline
Parameter & \multicolumn{3}{c}{Prior$\quad\;$} & \multicolumn{3}{c}{\db} & \multicolumn{3}{c}{\dbin} &           \\
          &                                  &&& & $s_{D,1}$ fix & $s_{D,0}$ \& $s_{D,1}$ fix   & \texttt{model 1} & \texttt{model 2} & \texttt{model 3} \\
\hline
$\gamma_p$                                                                 &                 2.2 &--& 2.5                  & $               {2.374}^{+0.004}_{-0.005}$ & $               {2.371}^{+0.004}_{-0.005}$ & $               {2.360}^{+0.005}_{-0.005}$ & $               {2.373}^{+0.005}_{-0.005}$ & $               {2.372}^{+0.005}_{-0.004}$ & $               {2.383}^{+0.005}_{-0.004}$ \\
$\gamma_{\rm{He}}-\gamma_p$                                                &                -0.2 &--& 0.1                  & $              {-0.062}^{+0.002}_{-0.002}$ & $              {-0.066}^{+0.002}_{-0.002}$ & $              {-0.063}^{+0.002}_{-0.002}$ & $              {-0.061}^{+0.002}_{-0.002}$ & $              {-0.062}^{+0.002}_{-0.002}$ & $              {-0.062}^{+0.002}_{-0.002}$ \\
$\gamma_{\rm{CNO}}-\gamma_p$                                               &                -0.2 &--& 0.1                  & $              {-0.025}^{+0.003}_{-0.003}$ & $              {-0.026}^{+0.003}_{-0.003}$ & $              {-0.029}^{+0.003}_{-0.003}$ & $              {-0.023}^{+0.003}_{-0.003}$ & $              {-0.024}^{+0.003}_{-0.003}$ & $              {-0.027}^{+0.003}_{-0.004}$ \\
$D_{0}\,\mathrm{[ 10^{28}\;cm^2/s]}$                                       &                 0.2 &--& 8.0                  & $                  {2.82}^{+0.29}_{-0.31}$ & $                  {4.16}^{+0.14}_{-0.11}$ & $                  {3.08}^{+0.09}_{-0.10}$ & $                  {0.52}^{+0.07}_{-0.07}$ & $                  {0.56}^{+0.05}_{-0.06}$ & $                  {2.46}^{+0.25}_{-0.21}$ \\
$\delta_{l}$                                                               &                -1.5 &--& 0.5                  & $                 {-0.80}^{+0.08}_{-0.06}$ & $                 {-0.77}^{+0.06}_{-0.05}$ & $                 {-0.42}^{+0.03}_{-0.03}$ & $                 {-0.78}^{+0.08}_{-0.06}$ & $                 {-0.79}^{+0.07}_{-0.07}$ & $                 {-0.84}^{+0.06}_{-0.06}$ \\
$\delta_{}$                                                                &                 0.1 &--& 2.0                  & $                  {0.72}^{+0.04}_{-0.12}$ & $               {0.500}^{+0.007}_{-0.007}$ & $               {0.585}^{+0.007}_{-0.008}$ & $                  {0.68}^{+0.04}_{-0.08}$ & $                  {0.68}^{+0.04}_{-0.09}$ & $                  {0.80}^{+0.06}_{-0.12}$ \\
$\delta_h-\delta$                                                          &                -2.0 &--& 1.0                  & $                 {-0.60}^{+0.22}_{-0.06}$ & $              {-0.189}^{+0.008}_{-0.009}$ & $              {-0.228}^{+0.008}_{-0.008}$ & $                 {-0.46}^{+0.10}_{-0.06}$ & $                 {-0.54}^{+0.14}_{-0.08}$ & $                 {-0.73}^{+0.25}_{-0.09}$ \\
$R_{0,D}\,\mathrm{[ 10^{3}\;MV]}$                                          &                 0.8 &--& 15.0                 & $                  {4.15}^{+0.26}_{-0.29}$ & $                  {3.63}^{+0.16}_{-0.21}$ & $                  {5.48}^{+0.14}_{-0.12}$ & $                  {4.27}^{+0.28}_{-0.24}$ & $                  {4.15}^{+0.25}_{-0.32}$ & $                  {4.23}^{+0.22}_{-0.24}$ \\
$s_{D,0}$                                                                  &                 0.1 &--& 1.0                  & $                  {0.30}^{+0.05}_{-0.05}$ & $                  {0.36}^{+0.02}_{-0.02}$ & --                                         & $                  {0.26}^{+0.04}_{-0.04}$ & $                  {0.31}^{+0.05}_{-0.04}$ & $                  {0.29}^{+0.03}_{-0.05}$ \\
$R_{D,1}\;\mathrm{[ 10^{3}\;MeV]}$                                         &                10.0 &--& 1000.0               & $              {285.67}^{+42.41}_{-97.61}$ & $              {216.99}^{+10.69}_{-14.82}$ & $                {172.17}^{+9.21}_{-9.93}$ & $              {189.85}^{+26.36}_{-22.47}$ & $              {290.20}^{+50.21}_{-80.82}$ & $              {257.05}^{+66.70}_{-72.19}$ \\
$s_{D,1}$                                                                  &                 0.1 &--& 5.0                  & $                  {1.42}^{+0.20}_{-0.56}$ & --                                         & --                                         & $                  {0.82}^{+0.12}_{-0.22}$ & $                  {1.26}^{+0.20}_{-0.41}$ & $                  {1.69}^{+0.25}_{-0.67}$ \\
$v_{0,\mathrm{c}}\,\mathrm{[km/s]}$                                        &                 0.0 &--& 40.0                 & $                 {13.10}^{+2.40}_{-2.29}$ & $                  {1.42}^{+0.38}_{-1.40}$ & $                 {13.13}^{+0.75}_{-0.67}$ & $                 {15.78}^{+3.13}_{-2.81}$ & $                 {14.68}^{+3.18}_{-2.33}$ & $                 {11.21}^{+1.33}_{-1.39}$ \\
Ren Abd$_p$                                                                &                 0.9 &--& 1.1                  & $               {1.010}^{+0.002}_{-0.002}$ & $               {1.014}^{+0.002}_{-0.002}$ & $               {1.006}^{+0.003}_{-0.003}$ & $               {1.004}^{+0.002}_{-0.002}$ & $               {1.009}^{+0.002}_{-0.002}$ & $               {1.011}^{+0.002}_{-0.002}$  & * \\
Ren Abd$_{^4\rm{He}}$                                                      &                 0.9 &--& 1.1                  & $               {1.001}^{+0.004}_{-0.004}$ & $               {1.012}^{+0.004}_{-0.004}$ & $               {1.006}^{+0.005}_{-0.005}$ & $               {0.994}^{+0.004}_{-0.004}$ & $               {1.001}^{+0.004}_{-0.004}$ & $               {0.995}^{+0.005}_{-0.004}$  & * \\
Abd$_{^{12}C}\,[ 10^{4} ] $                                                &                 0.2 &--& 0.5                  & $               {0.358}^{+0.005}_{-0.004}$ & $               {0.360}^{+0.004}_{-0.005}$ & $               {0.351}^{+0.004}_{-0.005}$ & $               {0.358}^{+0.005}_{-0.004}$ & $               {0.359}^{+0.004}_{-0.005}$ & $               {0.356}^{+0.005}_{-0.004}$ \\
Abd$_{^{14}N}\,[ 10^{4} ] $                                                &                 0.0 &--& 0.0                  & $               {0.024}^{+0.001}_{-0.002}$ & $               {0.024}^{+0.001}_{-0.002}$ & $               {0.024}^{+0.002}_{-0.002}$ & $               {0.024}^{+0.002}_{-0.002}$ & $               {0.024}^{+0.002}_{-0.001}$ & $               {0.023}^{+0.001}_{-0.002}$ \\
Abd$_{^{16}O}\,[ 10^{4} ] $                                                &                 0.2 &--& 0.6                  & $               {0.445}^{+0.003}_{-0.003}$ & $               {0.448}^{+0.003}_{-0.003}$ & $               {0.449}^{+0.003}_{-0.003}$ & $               {0.441}^{+0.003}_{-0.003}$ & $               {0.445}^{+0.003}_{-0.003}$ & $               {0.444}^{+0.004}_{-0.002}$ \\
$A_{\mathrm{pbaroverp,AMS-02}}$                                            &                 0.7 &--& 1.3                  & $                  {1.13}^{+0.01}_{-0.01}$ & $               {1.110}^{+0.013}_{-0.009}$ & $                  {1.13}^{+0.01}_{-0.01}$ & $                  {1.13}^{+0.02}_{-0.01}$ & $                  {1.13}^{+0.01}_{-0.01}$ & $               {1.147}^{+0.013}_{-0.009}$  & * \\
$A_\mathrm{XS}\;\; ^{4}\mathrm{He} \rightarrow ^{3}\mathrm{He} $           &                 0.7 &--& 1.3                  & $                  {1.27}^{+0.02}_{-0.01}$ & $                  {1.27}^{+0.02}_{-0.01}$ & $                  {1.27}^{+0.02}_{-0.01}$ & $                  {1.26}^{+0.02}_{-0.02}$ & $                  {1.26}^{+0.02}_{-0.01}$ & $               {1.275}^{+0.021}_{-0.007}$ \\
$A_\mathrm{XS}\,\rightarrow \mathrm{Li}$                                   &                 0.7 &--& 1.3                  & $               {1.295}^{+0.005}_{-0.000}$ & $               {1.295}^{+0.005}_{-0.000}$ & $               {1.294}^{+0.006}_{-0.000}$ & $               {1.293}^{+0.007}_{-0.001}$ & $               {1.294}^{+0.006}_{-0.000}$ & $              {1.295}^{+0.005}_{--0.000}$  & * \\
$A_\mathrm{XS}\,\rightarrow \mathrm{Be}$                                   &                 0.7 &--& 1.3                  & $               {1.045}^{+0.009}_{-0.005}$ & $               {1.049}^{+0.009}_{-0.005}$ & $               {1.048}^{+0.011}_{-0.004}$ & $               {1.033}^{+0.010}_{-0.006}$ & $               {1.044}^{+0.009}_{-0.005}$ & $               {1.010}^{+0.009}_{-0.006}$  & * \\
$A_\mathrm{XS}\,\rightarrow \mathrm{B}$                                    &                 0.7 &--& 1.3                  & $               {1.047}^{+0.010}_{-0.005}$ & $               {1.048}^{+0.009}_{-0.005}$ & $               {1.053}^{+0.010}_{-0.006}$ & $               {1.045}^{+0.010}_{-0.006}$ & $               {1.046}^{+0.009}_{-0.005}$ & $               {1.052}^{+0.010}_{-0.005}$  & * \\
$A_\mathrm{XS}\,\rightarrow \mathrm{C}$                                    &                 0.3 &--& 2.0                  & $                  {0.77}^{+0.11}_{-0.12}$ & $                  {0.79}^{+0.12}_{-0.11}$ & $                  {1.01}^{+0.12}_{-0.10}$ & $                  {0.71}^{+0.13}_{-0.11}$ & $                  {0.75}^{+0.11}_{-0.13}$ & $                  {0.81}^{+0.09}_{-0.14}$ \\
$A_\mathrm{XS}\,\rightarrow \mathrm{N}$                                    &                 0.7 &--& 1.3                  & $                  {1.13}^{+0.03}_{-0.03}$ & $                  {1.14}^{+0.03}_{-0.04}$ & $                  {1.16}^{+0.03}_{-0.04}$ & $                  {1.12}^{+0.03}_{-0.04}$ & $                  {1.13}^{+0.03}_{-0.04}$ & $                  {1.14}^{+0.04}_{-0.03}$ \\
$\delta_\mathrm{XS}\;\; ^{4}\mathrm{He} \rightarrow ^{3}\mathrm{He} $      &                -0.3 &--& 0.3                  & $                  {0.01}^{+0.02}_{-0.02}$ & $                  {0.03}^{+0.02}_{-0.02}$ & $                  {0.01}^{+0.02}_{-0.02}$ & $                  {0.00}^{+0.02}_{-0.02}$ & $                  {0.01}^{+0.02}_{-0.02}$ & $                  {0.01}^{+0.02}_{-0.02}$ \\
$\delta_\mathrm{XS}\,\rightarrow \mathrm{Li}$                              &                -0.3 &--& 0.3                  & $                  {0.11}^{+0.01}_{-0.01}$ & $                  {0.13}^{+0.01}_{-0.01}$ & $                  {0.09}^{+0.01}_{-0.01}$ & $                  {0.11}^{+0.01}_{-0.01}$ & $                  {0.11}^{+0.01}_{-0.01}$ & $                  {0.11}^{+0.01}_{-0.01}$ \\
$\delta_\mathrm{XS}\,\rightarrow \mathrm{Be}$                              &                -0.3 &--& 0.3                  & $                  {0.02}^{+0.02}_{-0.02}$ & $                  {0.03}^{+0.02}_{-0.02}$ & $                 {-0.01}^{+0.01}_{-0.02}$ & $                 {-0.01}^{+0.02}_{-0.02}$ & $                  {0.00}^{+0.02}_{-0.02}$ & $                  {0.03}^{+0.02}_{-0.01}$ \\
$\delta_\mathrm{XS}\,\rightarrow \mathrm{B}$                               &                -0.3 &--& 0.3                  & $                 {-0.04}^{+0.01}_{-0.01}$ & $                 {-0.02}^{+0.01}_{-0.01}$ & $                 {-0.05}^{+0.01}_{-0.01}$ & $                 {-0.04}^{+0.01}_{-0.01}$ & $                 {-0.04}^{+0.01}_{-0.01}$ & $              {-0.047}^{+0.016}_{-0.009}$ \\
$\delta_\mathrm{XS}\,\rightarrow \mathrm{C}$                               &                -0.3 &--& 0.3                  & $                  {0.24}^{+0.06}_{-0.02}$ & $                  {0.26}^{+0.04}_{-0.01}$ & $               {0.266}^{+0.034}_{-0.008}$ & $                  {0.24}^{+0.06}_{-0.02}$ & $                  {0.24}^{+0.06}_{-0.02}$ & $                  {0.24}^{+0.05}_{-0.02}$ \\
$\delta_\mathrm{XS}\,\rightarrow \mathrm{N}$                               &                -0.3 &--& 0.3                  & $                 {-0.03}^{+0.01}_{-0.02}$ & $                 {-0.01}^{+0.02}_{-0.01}$ & $                 {-0.03}^{+0.01}_{-0.02}$ & $                 {-0.03}^{+0.01}_{-0.02}$ & $                 {-0.03}^{+0.02}_{-0.02}$ & $              {-0.022}^{+0.015}_{-0.008}$ \\
$\Delta_{A_{B/C},{\rm DAMPE}}$                                                             &                -0.1 &--& 0.1                  & $               {0.050}^{+0.005}_{-0.005}$ & $               {0.051}^{+0.005}_{-0.004}$ & $               {0.050}^{+0.004}_{-0.005}$ & $               {0.048}^{+0.005}_{-0.004}$ & $               {0.051}^{+0.005}_{-0.004}$ & $               {0.040}^{+0.003}_{-0.004}$  & * \\
$\varphi_{\mathrm{AMS-02}}\,\mathrm{[GV]}$                                 &                 0.1 &--& 1.0                  & $                  {0.46}^{+0.01}_{-0.01}$ & $                  {0.48}^{+0.01}_{-0.01}$ & $               {0.397}^{+0.008}_{-0.009}$ & $                  {0.45}^{+0.01}_{-0.01}$ & $                  {0.46}^{+0.01}_{-0.01}$ & $                  {0.45}^{+0.01}_{-0.01}$  & * \\
$\varphi_{\mathrm{pbaroverp,AMS-02}}\,\mathrm{[GV]}$                       &                     &  &                      & $                  {0.74}^{+0.03}_{-0.03}$ & $                  {0.62}^{+0.03}_{-0.02}$ & $                  {0.71}^{+0.02}_{-0.02}$ & $                  {0.72}^{+0.03}_{-0.03}$ & $                  {0.75}^{+0.03}_{-0.03}$ & $                  {0.74}^{+0.02}_{-0.03}$ \\
$\chi^2_{\mathrm{p,AMS-02}}$ [72]                                              &                     &  &                      & $                                    24.0$ & $                                    65.2$ & $                                   109.4$ & $                                    27.1$ & $                                    25.1$ & $                                    22.9$ \\
$\chi^2_{\mathrm{p,CALET}}$ [23]                                              &                     &  &                      & $                                     5.0$ & $                                    16.0$ & $                                    25.3$ & $                                     5.2$ & $                                     5.2$ & $                                     4.8$ \\
$\chi^2_{\mathrm{p,Voyager}}$  [7]                                            &                     &  &                      & $                                     5.5$ & $                                     6.4$ & $                                    15.3$ & $                                     4.8$ & $                                     5.4$ & $                                     6.5$ \\
$\chi^2_{\mathrm{He,AMS-02}}$ [68]                                              &                     &  &                      & $                                     9.1$ & $                                    32.9$ & $                                    54.2$ & $                                    13.9$ & $                                     9.2$ & $                                     9.4$ \\
$\chi^2_{\mathrm{He,DAMPE}}$  [21]                                             &                     &  &                      & $                                     6.9$ & $                                    14.0$ & $                                    18.9$ & $                                     6.9$ & $                                     7.0$ & $                                     6.8$ \\
$\chi^2_{\mathrm{He,Voyager}}$  [8]                                           &                     &  &                      & $                                    20.1$ & $                                    28.4$ & $                                    22.7$ & $                                    16.3$ & $                                    19.4$ & $                                    21.8$ \\
$\chi^2_{\mathrm{C,AMS-02}}$ [68]                                              &                     &  &                      & $                                    45.3$ & $                                    40.4$ & $                                    50.4$ & $                                    47.7$ & $                                    44.4$ & $                                    48.2$ \\
$\chi^2_{\mathrm{N,AMS-02}}$ [66]                                             &                     &  &                      & $                                    49.3$ & $                                    50.1$ & $                                    49.3$ & $                                    45.9$ & $                                    48.7$ & $                                    49.1$ \\
$\chi^2_{\mathrm{O,AMS-02}}$  [67]                                             &                     &  &                      & $                                    22.1$ & $                                    26.8$ & $                                    47.3$ & $                                    20.7$ & $                                    21.4$ & $                                    25.0$ \\
$\chi^2_{\mathrm{3heover4he,AMS-02}}$  [26]                                    &                     &  &                      & $                                     5.5$ & $                                     7.3$ & $                                     4.3$ & $                                     5.5$ & $                                     5.6$ & $                                     5.4$ \\
$\chi^2_{\mathrm{Li/C,AMS-02}}$  [67]                                          &                     &  &                      & $                                    55.5$ & $                                    62.3$ & $                                    49.2$ & $                                    50.8$ & $                                    56.3$ & $                                    50.9$ \\
$\chi^2_{\mathrm{Be/C,AMS-02}}$ [67]                                           &                     &  &                      & $                                    33.8$ & $                                    30.8$ & $                                    33.1$ & $                                    32.3$ & $                                    30.6$ & $                                    60.3$ \\
$\chi^2_{\mathrm{B/C,AMS-02}}$ [67]                                            &                     &  &                      & $                                    33.5$ & $                                    37.3$ & $                                    43.2$ & $                                    31.9$ & $                                    31.4$ & $                                    47.2$ \\
$\chi^2_{\mathrm{pbaroverp,AMS-02}}$ [58]                                      &                     &  &                      & $                                    70.7$ & $                                    37.9$ & $                                    50.4$ & $                                    69.6$ & $                                    68.8$ & $                                    73.5$ \\
$\chi^2_{\rm{tot}}$ [685]                                                                    &                     &  &                      & $                                   385$ & $                                   458$ & $                                   582$ & $                                   377$ & $                                   378$ & $                                   427$ \\
\hline
    \end{tabular}
\end{table*}

\begin{table*}
    \caption{Same as the previous table but for the models that contain reacceleration.}
    \label{tab:full_results_injb}
    \begin{tabular}{l rcl c c c c c c }
\hline \hline
Parameter & \multicolumn{3}{c}{Prior$\quad\;$} & \multicolumn{3}{c}{\ib}           \\
          &                                  && & $s_{D,1}$ fix & $s_{D,0}$ \& $s_{D,1}$ fix   \\
\hline
$\gamma_{1,p}$                                                             &                 1.5 &--& 2.1                  & $                  {1.74}^{+0.02}_{-0.02}$ & $                  {1.75}^{+0.02}_{-0.02}$ & $               {1.853}^{+0.010}_{-0.009}$ \\
$\gamma_p$                                                                 &                 2.3 &--& 2.6                  & $               {2.422}^{+0.005}_{-0.004}$ & $               {2.420}^{+0.005}_{-0.004}$ & $               {2.411}^{+0.005}_{-0.005}$ \\
$\gamma_{1,^{4}\mathrm{He}}$                                               &                 1.5 &--& 2.1                  & $                  {1.79}^{+0.02}_{-0.02}$ & $                  {1.80}^{+0.02}_{-0.02}$ & $                  {1.90}^{+0.01}_{-0.01}$ \\
$\gamma_{2,^{4}\mathrm{He}}$                                               &                 2.2 &--& 2.5                  & $               {2.360}^{+0.005}_{-0.004}$ & $               {2.359}^{+0.005}_{-0.004}$ & $               {2.352}^{+0.004}_{-0.004}$ \\
$\gamma_{1}$                                                               &                 1.8 &--& 2.3                  & $                  {2.07}^{+0.02}_{-0.02}$ & $                  {2.07}^{+0.02}_{-0.02}$ & $                  {2.15}^{+0.01}_{-0.02}$ \\
$\gamma_{2}$                                                               &                 2.2 &--& 2.6                  & $               {2.399}^{+0.004}_{-0.004}$ & $               {2.399}^{+0.005}_{-0.004}$ & $               {2.396}^{+0.005}_{-0.004}$ \\
$R_{\rm{inj}}\;\mathrm{[ 10^{3}\;MeV]}$                                    &                 2.0 &--& 25.0                 & $                  {7.58}^{+0.39}_{-0.36}$ & $                  {7.69}^{+0.39}_{-0.37}$ & $                  {9.92}^{+0.32}_{-0.33}$ \\
$s_{inj}$                                                                  &                 0.1 &--& 1.0                  & $                  {0.33}^{+0.03}_{-0.02}$ & $                  {0.31}^{+0.02}_{-0.03}$ & --                                         \\
$D_{0}\,\mathrm{[ 10^{28}\;cm^2/s]}$                                       &                 1.0 &--& 10.0                 & $                  {4.53}^{+0.07}_{-0.10}$ & $                  {4.57}^{+0.09}_{-0.09}$ & $                  {4.62}^{+0.22}_{-0.14}$ \\
$\delta_{}$                                                                &                 0.1 &--& 1.0                  & $               {0.406}^{+0.005}_{-0.005}$ & $               {0.402}^{+0.005}_{-0.005}$ & $               {0.393}^{+0.007}_{-0.006}$ \\
$\delta_h-\delta$                                                          &                -1.0 &--& 0.1                  & $                 {-0.21}^{+0.02}_{-0.02}$ & $              {-0.164}^{+0.010}_{-0.010}$ & $                 {-0.16}^{+0.01}_{-0.01}$ \\
$R_{D,1}\;\mathrm{[ 10^{3}\;MeV]}$                                         &               100.0 &--& 900.0                & $              {390.30}^{+40.35}_{-59.30}$ & $              {298.40}^{+22.59}_{-23.88}$ & $              {367.57}^{+29.01}_{-36.20}$ \\
$s_{D,1}$                                                                  &                 0.1 &--& 1.0                  & $                  {0.35}^{+0.08}_{-0.08}$ & --                                         & --                                         \\
$v_{\mathrm{A}}\,\mathrm{[km/s]}$                                          &                 0.0 &--& 60.0                 & $                 {23.82}^{+0.72}_{-0.69}$ & $                 {24.12}^{+0.74}_{-0.72}$ & $                 {25.63}^{+1.05}_{-0.98}$ \\
Ren Abd$_p$                                                                &                 0.9 &--& 1.1                  & $               {1.003}^{+0.002}_{-0.002}$ & $               {1.005}^{+0.002}_{-0.002}$ & $               {1.007}^{+0.002}_{-0.001}$  & * \\
Ren Abd$_{^4\rm{He}}$                                                      &                 0.9 &--& 1.1                  & $               {1.015}^{+0.005}_{-0.004}$ & $               {0.995}^{+0.004}_{-0.004}$ & $               {1.004}^{+0.004}_{-0.004}$  & * \\
Abd$_{^{12}C}\,[ 10^{4} ] $                                                &                 0.2 &--& 0.5                  & $               {0.357}^{+0.003}_{-0.002}$ & $               {0.358}^{+0.003}_{-0.002}$ & $               {0.360}^{+0.004}_{-0.003}$ \\
Abd$_{^{14}N}\,[ 10^{4} ] $                                                &                 0.0 &--& 0.0                  & $               {0.023}^{+0.002}_{-0.001}$ & $               {0.024}^{+0.002}_{-0.002}$ & $               {0.024}^{+0.002}_{-0.001}$ \\
Abd$_{^{16}O}\,[ 10^{4} ] $                                                &                 0.2 &--& 0.6                  & $               {0.426}^{+0.003}_{-0.003}$ & $               {0.427}^{+0.003}_{-0.003}$ & $               {0.432}^{+0.003}_{-0.004}$ \\
$A_{\mathrm{pbaroverp,AMS-02}}$                                            &                 0.7 &--& 1.3                  & $               {1.133}^{+0.015}_{-0.010}$ & $               {1.123}^{+0.015}_{-0.009}$ & $                  {1.08}^{+0.04}_{-0.02}$  & * \\
$A_\mathrm{XS}\;\; ^{4}\mathrm{He} \rightarrow ^{3}\mathrm{He} $           &                 0.7 &--& 1.3                  & $                  {1.22}^{+0.02}_{-0.02}$ & $                  {1.22}^{+0.02}_{-0.02}$ & $                  {1.17}^{+0.04}_{-0.03}$ \\
$A_\mathrm{XS}\,\rightarrow \mathrm{Li}$                                   &                 0.7 &--& 1.3                  & $               {1.294}^{+0.006}_{-0.001}$ & $               {1.294}^{+0.006}_{-0.001}$ & $                  {1.26}^{+0.04}_{-0.01}$  & * \\
$A_\mathrm{XS}\,\rightarrow \mathrm{Be}$                                   &                 0.7 &--& 1.3                  & $               {1.041}^{+0.009}_{-0.005}$ & $               {1.041}^{+0.010}_{-0.004}$ & $                  {1.02}^{+0.03}_{-0.02}$  & * \\
$A_\mathrm{XS}\,\rightarrow \mathrm{B}$                                    &                 0.7 &--& 1.3                  & $               {1.036}^{+0.008}_{-0.005}$ & $               {1.035}^{+0.008}_{-0.005}$ & $                  {1.01}^{+0.03}_{-0.01}$  & * \\
$A_\mathrm{XS}\,\rightarrow \mathrm{C}$                                    &                 0.3 &--& 2.0                  & $                  {0.37}^{+0.02}_{-0.07}$ & $                  {0.37}^{+0.02}_{-0.07}$ & $                  {0.39}^{+0.03}_{-0.09}$ \\
$A_\mathrm{XS}\,\rightarrow \mathrm{N}$                                    &                 0.7 &--& 1.3                  & $                  {1.12}^{+0.04}_{-0.04}$ & $                  {1.11}^{+0.03}_{-0.04}$ & $                  {1.08}^{+0.04}_{-0.04}$ \\
$\delta_\mathrm{XS}\;\; ^{4}\mathrm{He} \rightarrow ^{3}\mathrm{He} $      &                -0.3 &--& 0.3                  & $                  {0.25}^{+0.02}_{-0.01}$ & $                  {0.25}^{+0.01}_{-0.02}$ & $                  {0.24}^{+0.02}_{-0.01}$ \\
$\delta_\mathrm{XS}\,\rightarrow \mathrm{Li}$                              &                -0.3 &--& 0.3                  & $               {0.290}^{+0.010}_{-0.002}$ & $               {0.287}^{+0.013}_{-0.003}$ & $               {0.281}^{+0.016}_{-0.008}$ \\
$\delta_\mathrm{XS}\,\rightarrow \mathrm{Be}$                              &                -0.3 &--& 0.3                  & $                  {0.21}^{+0.01}_{-0.02}$ & $                  {0.20}^{+0.01}_{-0.01}$ & $                  {0.19}^{+0.01}_{-0.02}$ \\
$\delta_\mathrm{XS}\,\rightarrow \mathrm{B}$                               &                -0.3 &--& 0.3                  & $               {0.131}^{+0.011}_{-0.010}$ & $               {0.128}^{+0.010}_{-0.010}$ & $                  {0.13}^{+0.01}_{-0.01}$ \\
$\delta_\mathrm{XS}\,\rightarrow \mathrm{C}$                               &                -0.3 &--& 0.3                  & $                  {0.23}^{+0.07}_{-0.02}$ & $                  {0.23}^{+0.07}_{-0.02}$ & $                  {0.24}^{+0.06}_{-0.01}$ \\
$\delta_\mathrm{XS}\,\rightarrow \mathrm{N}$                               &                -0.3 &--& 0.3                  & $                  {0.09}^{+0.02}_{-0.02}$ & $                  {0.09}^{+0.02}_{-0.01}$ & $                  {0.08}^{+0.02}_{-0.02}$ \\
$\Delta_{A_{B/C},{\rm DAMPE}}$                                             &                -0.1 &--& 0.1                  & $               {0.033}^{+0.004}_{-0.004}$ & $               {0.034}^{+0.005}_{-0.004}$ & $               {0.036}^{+0.005}_{-0.005}$  & * \\
$\varphi_{\mathrm{AMS-02}}\,\mathrm{[GV]}$                                 &                 0.1 &--& 1.0                  & $               {0.656}^{+0.009}_{-0.011}$ & $               {0.655}^{+0.011}_{-0.009}$ & $               {0.676}^{+0.009}_{-0.009}$  & * \\
$\varphi_{\mathrm{pbaroverp,AMS-02}}\,\mathrm{[GV]}$                       &                     &  &                      & $                  {0.53}^{+0.03}_{-0.03}$ & $                  {0.51}^{+0.03}_{-0.02}$ & $                  {0.43}^{+0.04}_{-0.03}$ \\
$\chi^2_{\mathrm{p,AMS-02}}$                                               &                     &  &                      & $                                    22.1$ & $                                    27.0$ & $                                    92.3$ \\
$\chi^2_{\mathrm{p,CALET}}$                                                &                     &  &                      & $                                     6.4$ & $                                    10.8$ & $                                     9.2$ \\
$\chi^2_{\mathrm{p,Voyager}}$                                              &                     &  &                      & $                                    35.4$ & $                                    34.6$ & $                                    20.2$ \\
$\chi^2_{\mathrm{He,AMS-02}}$                                              &                     &  &                      & $                                    24.2$ & $                                    27.3$ & $                                    46.8$ \\
$\chi^2_{\mathrm{He,DAMPE}}$                                               &                     &  &                      & $                                     8.2$ & $                                    11.2$ & $                                     9.8$ \\
$\chi^2_{\mathrm{He,Voyager}}$                                             &                     &  &                      & $                                    30.7$ & $                                    29.7$ & $                                    43.9$ \\
$\chi^2_{\mathrm{C,AMS-02}}$                                               &                     &  &                      & $                                    26.5$ & $                                    25.8$ & $                                    32.3$ \\
$\chi^2_{\mathrm{N,AMS-02}}$                                               &                     &  &                      & $                                    34.9$ & $                                    36.5$ & $                                    41.9$ \\
$\chi^2_{\mathrm{O,AMS-02}}$                                               &                     &  &                      & $                                    12.6$ & $                                    13.7$ & $                                    19.0$ \\
$\chi^2_{\mathrm{3heover4he,AMS-02}}$                                      &                     &  &                      & $                                     7.3$ & $                                     7.3$ & $                                    15.5$ \\
$\chi^2_{\mathrm{Li/C,AMS-02}}$                                            &                     &  &                      & $                                    37.1$ & $                                    36.6$ & $                                    34.9$ \\
$\chi^2_{\mathrm{Be/C,AMS-02}}$                                            &                     &  &                      & $                                    41.7$ & $                                    40.4$ & $                                    45.8$ \\
$\chi^2_{\mathrm{B/C,AMS-02}}$                                             &                     &  &                      & $                                    35.0$ & $                                    35.0$ & $                                    34.9$ \\
$\chi^2_{\mathrm{pbaroverp,AMS-02}}$                                       &                     &  &                      & $                                    50.9$ & $                                    48.6$ & $                                    46.5$ \\
$\chi_{\rm{tot}}^2$                                                                    &                     &  &                      & $                                   359.0$ & $                                   371$ & $                                   482$ \\
\hline
    \end{tabular}
\end{table*}

\end{document}